\documentclass[11pt, a4paper, apj]{emulateapj}
\usepackage{epsfig}
\usepackage{amsmath}

\slugcomment{Not to appear in Nonlearned J., 45.}

\shorttitle{Structures of Bright-rimmed Clouds}
\shortauthors{K. Motoyama et al.}

\begin{document}

\title{EFFECTS OF MAGNETIC FIELD AND FUV RADIATION ON THE STRUCTURES OF BRIGHT-RIMMED CLOUDS}

\author{Kazutaka Motoyama}
\affil{National Institute of Informatics, 2-1-2 Hitotsubashi, Chiyoda-ku, Tokyo, Japan 101-8430 \email{motoyama@nii.ac.jp}}

\author{Tomofumi Umemoto} 
\affil{National Astronomical Observatory of Japan, 2-21-1 Osawa, Mitaka, Tokyo, Japan 191-8588}

\author{Hsien Shang\altaffilmark{1} and Tatsuhiko Hasegawa}
\affil{Institute of Astronomy and Astrophysics, Academia Sinica, Taipei, Taiwan 106}

\altaffiltext{1}{Theoretical Institute for Advanced Research in Astrophysics, National Tsing Hua University, 
101, Sec. 2, Kuang-Fu Rd., Hsin-Chu, Taiwan 30013}

\begin{abstract}
The bright-rimmed cloud SFO 22 was observed with the 45 m telescope of Nobeyama Radio Observatory in the 
$^{12}$CO (J = 1-0), $^{13}$CO (J = 1-0), and C$^{18}$O (J = 1-0) lines, where well-developed head-tail structure
and small line widths were found.  Such features were predicted by radiation-driven implosion models, suggesting that SFO 22 may be in a quasi-stationary equilibrium state. We compare the observed properties with those from numerical models of a photo-evaporating cloud, which include effects of magnetic pressure and heating due to strong far-ultraviolet (FUV) radiation from an exciting star. 
The magnetic pressure may play a more important role in the density structures of bright-rimmed clouds, than the thermal pressure that is enhanced by the FUV radiation. The FUV radiation can heat the cloud surface to near 30 K, however, its effect is not enough to reproduce the observed density structure of SFO 22.  An initial magnetic field of 5 $\mathrm{\mu G}$ in our numerical models produces the best agreement with the observations, and its direction can affect the structures of bright-rimmed clouds.
\end{abstract}


\section{Introduction}

Bright-rimmed clouds (BRCs) are cometary molecular clouds found at the edge of \ion{H}{2} regions.
These clouds have bright rims on the side facing the exciting star and extended tails on the other side. Since their 
head-tail morphologies suggest that BRCs are interacting with the radiation or stellar wind from an exciting star, BRCs are considered as 
potential sites of triggered star formation. The gradients of age spread in
young stars along the axes of BRCs indicate that their formation may have been
sequentially triggered by shock waves
\citep{1995ApJ...455L..39S,2007ApJ...654..316G,2008AJ....135.2323I,2009ApJ...699.1454G}.  

Radiation-driven implosion models are often considered for the formation and evolution of BRCs and their triggered origins. 
Strong UV radiation from nearby massive stars can photoionize and photoevaporate the surfaces of surrounding molecular clouds, and whose effects have been studied by various groups. \citet{1989ApJ...346..735B} developed an approximate analytical solution for the evolution of molecular cloud compressed by radiation-driven implosion. \citet{1994A&A...289..559L} investigated a radiation-driven implosion model using hydrodynamic simulations. Recent hydrodynamic simulations of radiation-driven implosion include effects of physics such as self-gravity of the gas \citep{2003MNRAS.338..545K,2006MNRAS.369..143M}, diffuse radiation field \citep{2012MNRAS.420..562H}, and turbulence in molecular clouds \citep{2009ApJ...694L..26G}. \citet{2007A&A...467..657M} demonstrated that radiation-driven implosion can enhance accretion rates enough to account for the high luminosities of YSOs observed in the BRCs \citep{1989ApJ...342L..87S}. 
Typical shock speed of a few $\mathrm{km \ s^{-1}}$ in this study is consistent with that 
estimated from observations of age gradients of young stars in and aroud BRCs 
\citep{2007ApJ...654..316G,2009ApJ...699.1454G}.

Radiation-MHD studies suggest the possibility of altering evolution of photoionized cloud by the presence of magnetic field.
\cite{2009MNRAS.398..157H} carried out the first three-dimensional radiation-MHD simulations of photoionization of a magnetized molecular globule under ultraviolet radiation. They reported the photo-evaporating globule will evolve into a more flatten shape compared with the non-magnetic case when the cloud initially has a strong magnetic field (that is 100 times the thermal pressure) perpendicular to UV radiation field.  \cite{2011MNRAS.412.2079M} also showed that strong magnetic field has significant influence on the dynamics of the photoionization process. Measuring magnetic field strengths in BRCs observationally has been difficult, and the effects of magnetic field can only be inferred indirectly by comparing the observed density structures and kinematics with those obtained by theoretical models.

Strong far-ultraviolet (FUV) radiation from an exciting star may also influence the evolution of a photoionized cloud. 
As shown in many studies of photon dominated regions \citep[e.g.][]{Tielens1985,Hollenbach1991}, the FUV radiation from a massive star can heat molecular clouds through photoelectric heating and photodissociation of important coolants such as carbon monoxide. Temperature of the cloud heated by FUV radiation ranges from a few tens up to a few hundreds of Kelvin, strongly dependent
on the intensity of the FUV radiation and the density of cloud. High thermal pressure enhanced by FUV radiation may affect the evolution of BRCs. However, there have been limited theoretical works that include heating due to FUV radiation \citep{2006MNRAS.369..143M,2009MNRAS.398..157H}. These effects should be included in theoretical models for reliable comparisons with observations.

The purpose of this study is to investigate how magnetic field and FUV radiation affect the evolution of BRCs through the comparisons between observations and numerical models.  An evolutionary scenario of BRCs due to radiation-driven implosion has an initial implosion phase followed by a quasi-stationary equilibrium phase. In this paper, we focus on the quasi-stationary equilibrium phase, and implosion phase will be investigated in a subsequent paper. The BRC SFO 22 was observed with the 45 m telescope of Nobeyama Radio Observatory, and the results were compared with numerical models of photoionized clouds with effects of magnetic field and FUV radiation. The layout of this paper is as follows. In section \ref{sect observations}, the details of observations are described.
In section \ref{sect model}, a description of our numerical models is given. Section \ref{sect results} and section \ref{sect discussion} give the results and discussions. In section \ref{sect conclusions}, we summarize our main conclusions.


\section{Observations and Analysis}
\label{sect observations}

\subsection{Observed Bright-rimmed Cloud}

\begin{figure}
  \begin{center}
    \epsscale{1.0}
    \plotone{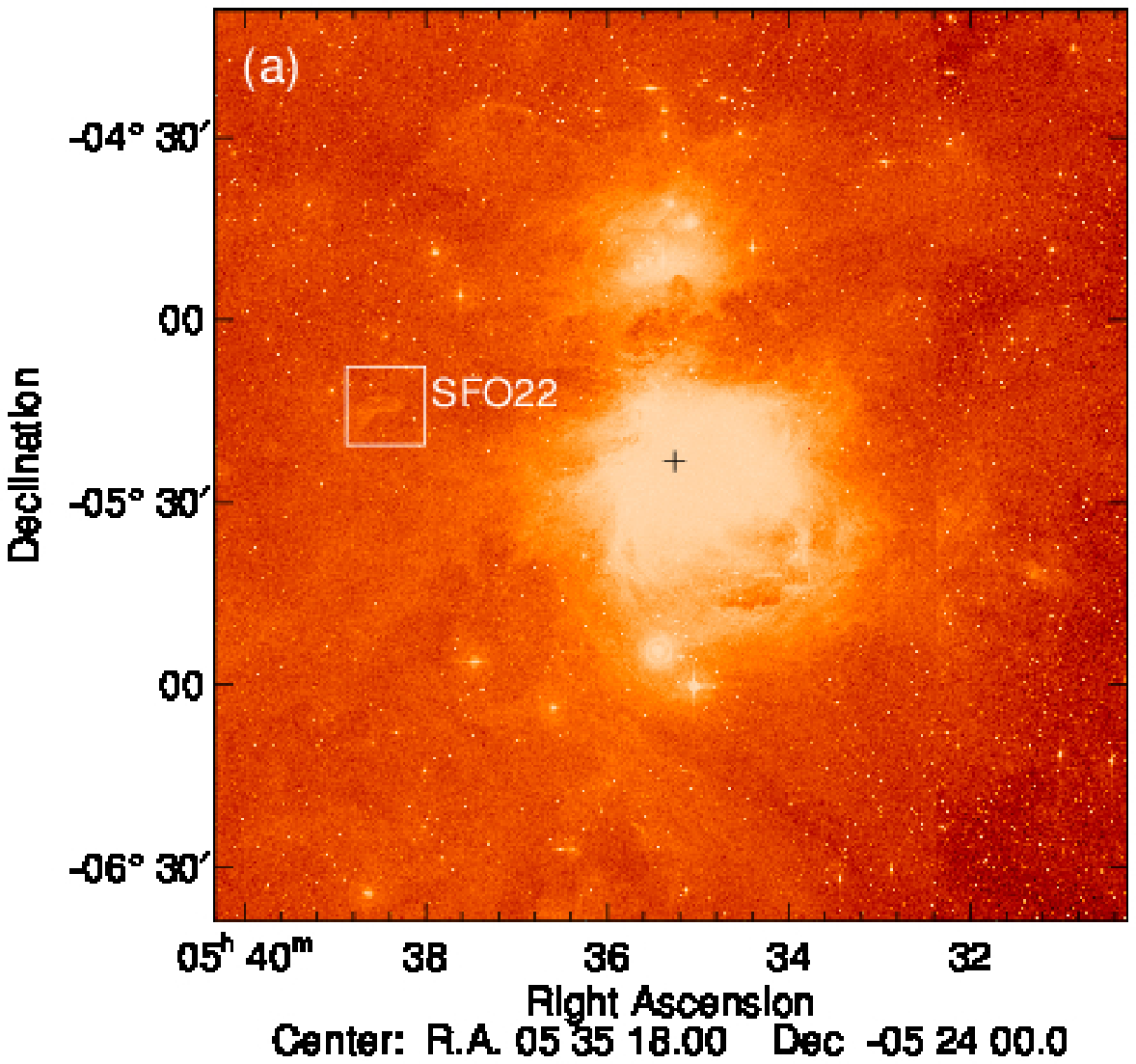}
    \epsscale{0.8}
    \plotone{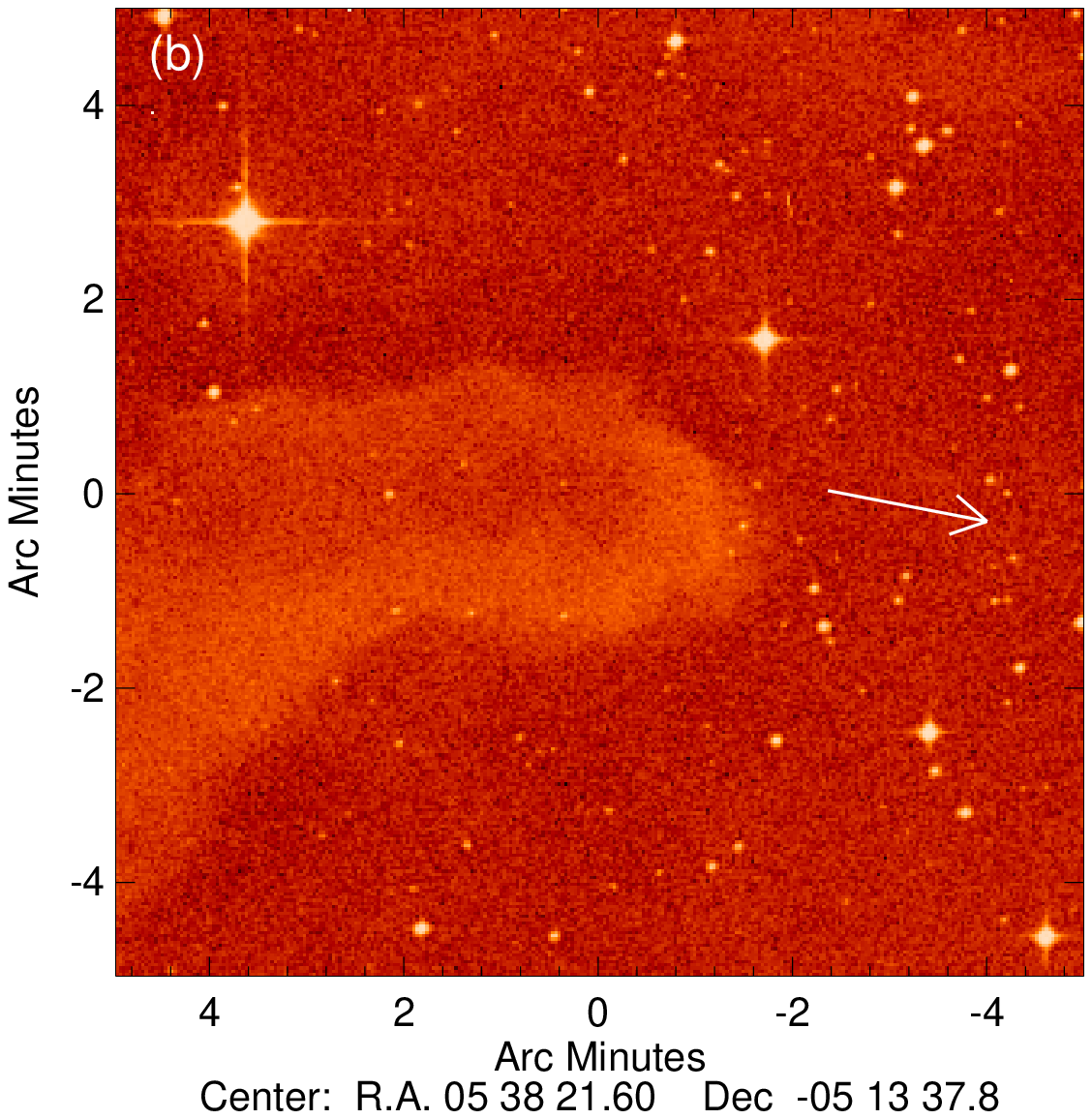}
  \end{center}
       \caption{(a) DSS red image of the \ion{H}{2} region s281. The cross labels the
     position of the exciting star. 
     (b) DSS red image of bright-rimmed cloud SFO 22. The arrow indicates the direction to the exciting star. \newline 
     (A color version of this figure is available in the online journal.)}
     \label{DSS images}
\end{figure}

BRC SFO 22 is located at the eastern edge of \ion{H}{2} region s281, and was selected from BRC catalog of \citet{1991ApJS...77...59S}.
Figure \ref{DSS images} (a) shows the entire image of the region s281, and \ref{DSS images} (b) gives the close-up view of SFO 22 enlarged from \ref{DSS images} (a). This \ion{H}{2} region is ionized by $\theta^1$Ori, marked by a cross in Figure \ref{DSS images} (a). 
The spectral type of the primary exciting star in $\theta^1$Ori is O7V, and has a projected distance of 6.5 pc to SFO 22 \citep{2004A&A...426..535M}. S281 \citep{1964ARA&A...2..213B} is 460 pc away from us.

\citet{1991ApJS...77...59S} classified the BRCs into three types in order of increasing degree of rim curvature: type A, B, and C. 
Based on the radiation-driven implosion model, the shapes of cloud rims reflect the evolutionary stages.  The type A BRCs are still undergoing compression by the shockwaves generated by ionization, and type B and type C BRCs are approaching or have already reached the phase of quasi-stationary equilibria. SFO 22 has a well-developed head-tail structure along the line to the exciting star and it is classified as type B.  Although the head part of the cloud contains the IRAS point source 05359-0515, ammonia rotational inversion lines were not detected toward this IRAS point source \citep{2010MNRAS.408..157M}. Since ammonia lines trace dense gas associated with protostellar cores, the non-detection of ammonia lines has been interpreted as no star forming activities in SFO 22. 

\subsection{Observations}

Observations were carried out using the 45 m telescope of Nobeyama Radio Observatory 
in 2005 January and April. We observed $^{12}$CO (J = 1-0) at 115.271203 GHz, 
$^{13}$CO (J = 1-0) at 110.201370 GHz, and C$^{18}$O (J = 1-0) at 109.782182 GHz. 
The half-power beam width of the telescope and main-beam efficiency at 115 GHz were 15\arcsec and $\eta_{MB}=0.4$, respectively.  
We used the 5$\times$5 beam focal plane array receiver "BEARS" whose beam separation is 41.1\arcsec.
As receiver backends we used a 1024 channel digital autocorrelator with a 31.25 kHz frequency resolution. 
The corresponding velocity resolution is 81.5 $\mathrm{m \, s^{-1}}$ at 115 GHz. The typical system noise temperature was 300-450 K depending on the atmospheric conditions. The intensity scale of the spectra was calibrated by the chopper wheel method. 
The corrected antenna temperature $T^*_A$ is converted into main beam brightness temperature using the relation of $T_B = T^*_A /\eta_{mb}$.

We observed SFO 22 with a grid spacing of 20.55\arcsec, which is half of the beam separation of BEARS, in the line of $^{12}$CO (J = 1-0).  The mapped area was $390\arcsec \times 390\arcsec$. Dense regions of clouds were observed with finer grid spacing of 10.3\arcsec in the lines of $^{13}$CO (J = 1-0) and C$^{18}$O (J = 1-0). 
The mapped area for these lines were $195\arcsec \times 195\arcsec$ and $154\arcsec \times 154\arcsec$, respectively. 
The pointing accuracy of the antenna was checked and corrected every 1.5-2 hr using SiO maser emission from Ori KL, and its typical error was less than 5\arcsec. All observations were carried out by position switching mode. 
The data were reduced by using the software package NewStar provided by Nobeyama Radio Observatory.


\subsection{Observational Results and Analysis}
 
\begin{figure}
  \begin{center} 

  \epsscale{0.95}
  \plotone{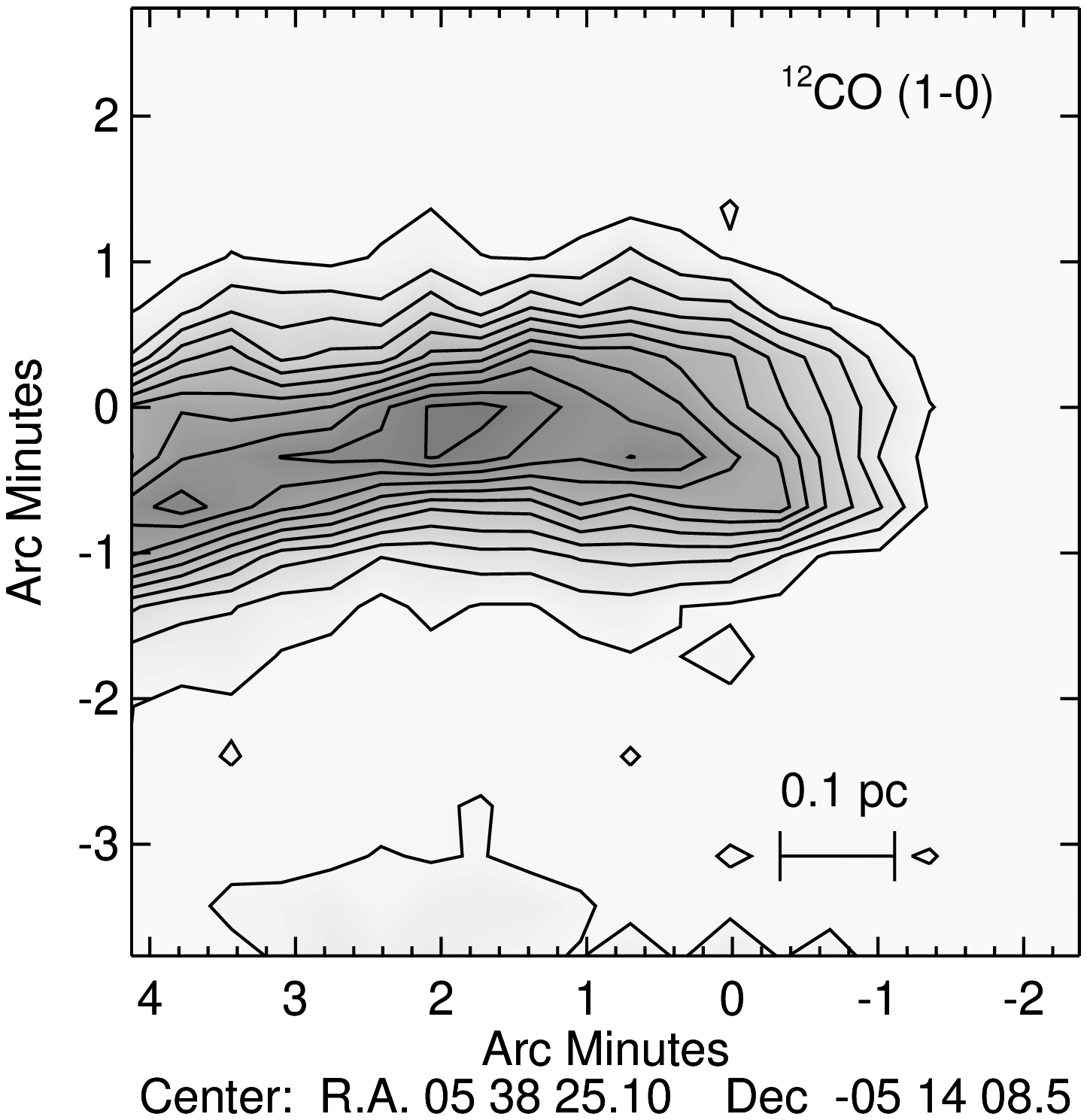}\\
  \plotone{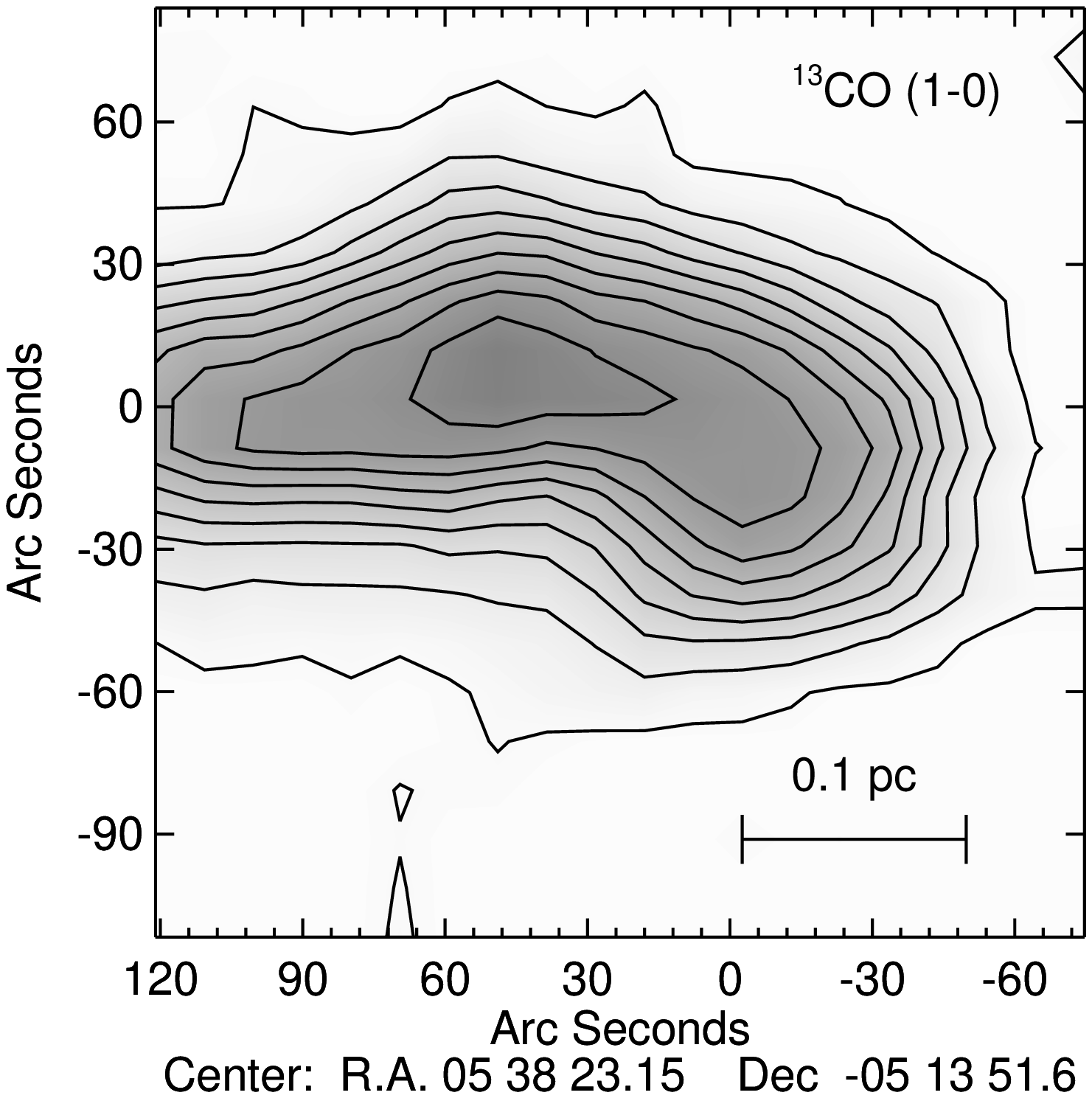}\\
  \plotone{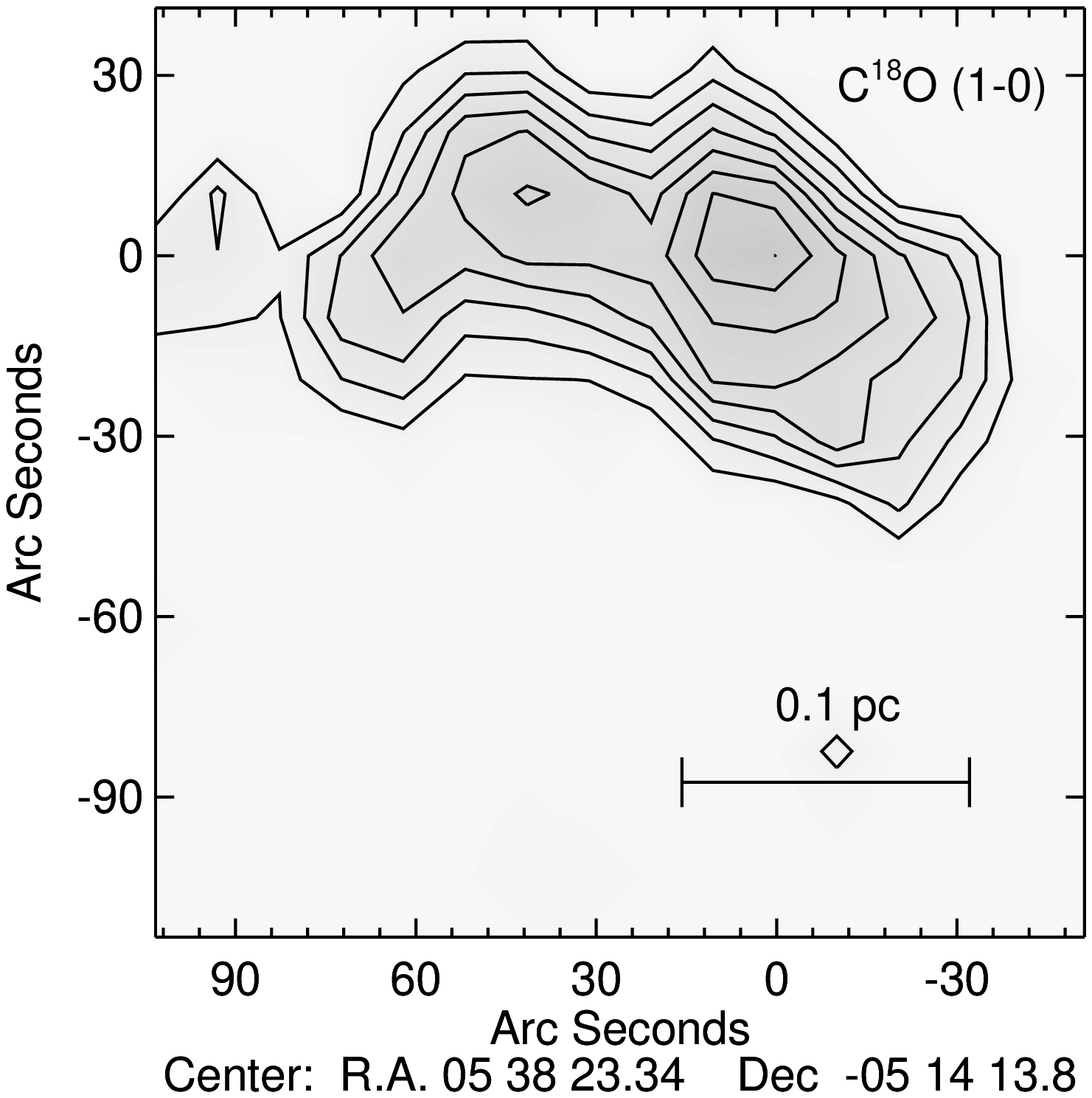}

  \caption{Integrated intensity maps of SFO 22. 
      The $^{12}$CO (J = 1-0) map (top) has a lowest contour of 1.39 K km ${\rm s^{-1}}$ (3 $\sigma$) and 
      contour intervals of 1.39 K km ${\rm s^{-1}}$ (3 $\sigma$).
      The $^{13}$CO (J = 1-0) map (middle) has a lowest contour of 0.315 K km ${\rm s^{-1}}$ (3 $\sigma$) and 
      contour intervals of 0.525 K km ${\rm s^{-1}}$ (5 $\sigma$).
      The C$^{18}$O (J = 1-0) map (bottom) has a lowest contour of 0.131 K km ${\rm s^{-1}}$ (3 $\sigma$) and 
      contour intervals of 0.0435 K km ${\rm s^{-1}}$ (1 $\sigma$).} 
  \label{intensity map sfo22}
  \end{center}
 \end{figure}

Figure \ref{intensity map sfo22} shows the velocity-integrated intensity maps of SFO 22.
The reference center of the map is the peak position of C$^{18}$O (J = 1-0) emission 
at RA(2000) =  5$\mathrm{^h}$ 38$\mathrm{^m}$ 21.6$\mathrm{^s}$, Dec(2000) = -5$\arcdeg$ 
13$\arcmin$ 37.8$\arcsec$. Emission from $^{12}$CO (J = 1-0) coincides with the optical images. 
The cometary morphology is clearly shown. On the contrary, the C$^{18}$O (J = 1-0) emission is very weak and detected only at a few points.

\begin{figure*}
  \begin{center} 
 
      \epsscale{1.0}
      \plotone{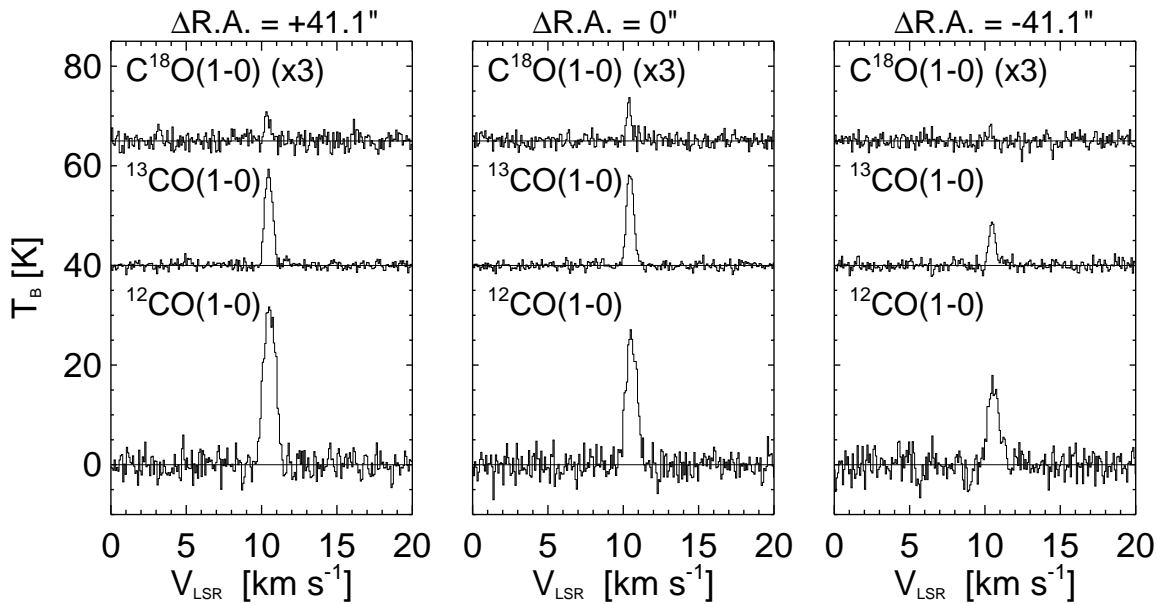} 

  \caption{Some selected $^{12}$CO (J = 1-0), $^{13}$CO (J = 1-0), and C$^{18}$O (J = 1-0) 
spectra observed toward the SFO 22 along declination through the peak of C$^{18}$O (J = 1-0). }

 \label{spectra sfo22} 
  \end{center}
 \end{figure*}

\begin{figure*}
  \begin{center} 

  \plotone{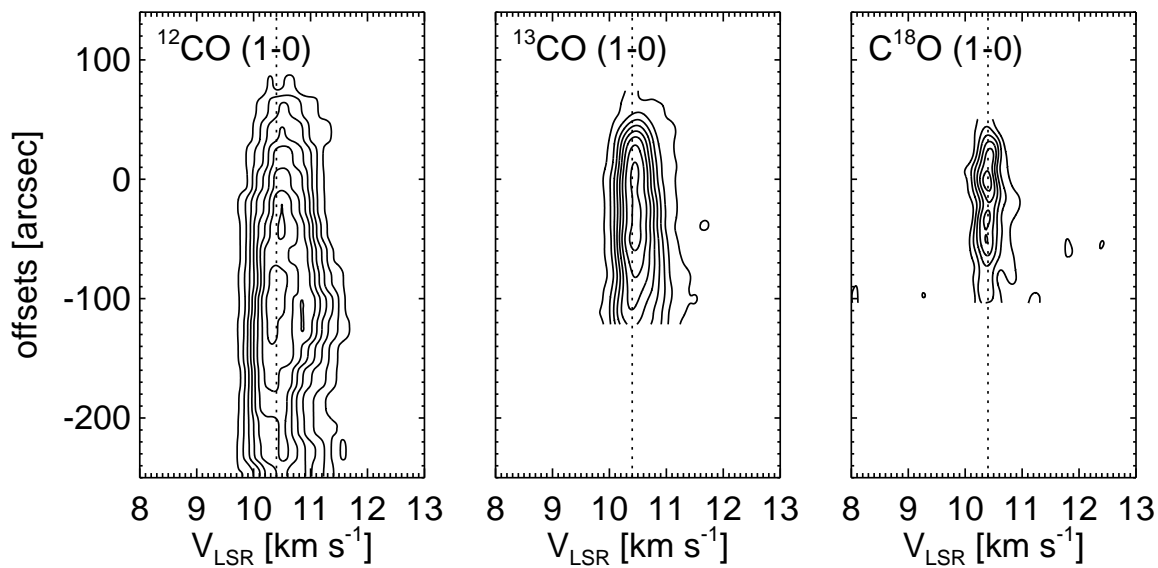}

  \caption{Position-velocity diagrams of $^{12}$CO (J = 1-0) (left), $^{13}$CO (J = 1-0) (center), 
      and C$^{18}$O (J = 1-0) (right) 
      along declination through the peak positions of C$^{18}$O (J = 1-0) emission.
      The diagram for $^{12}$CO (J = 1-0) has lowest contour of 1.53 K (3 $\sigma$) and 
      contour interval of 1.53 K  (3 $\sigma$).
      The diagram for $^{13}$CO (J = 1-0) has lowest contour of 0.276 K (3 $\sigma$) and 
      contour interval of 0.92 K (10 $\sigma$).
      The diagram for C$^{18}$O (J = 1-0) has lowest contour of 0.125 K (3 $\sigma$) and 
      contour interval of 0.125 K km (3 $\sigma$). The vertical dotted lines indicate the systemic velocity of 
      10.4 km/s.} 
 \label{pv map sfo22}
  \end{center}
 \end{figure*}

Figure \ref{spectra sfo22} shows the observed line spectra toward the peak position of 
C$^{18}$O (J = 1-0) and offset positions along declination. 
Assuming that $^{12}$CO (J = 1-0) emission is optically thick at the peak of 
C$^{18}$O (J = 1-0) emission, we can calculate the excitation temperature as
\begin{equation}
 T_{ex} = \frac{5.53}{\ln [1 + 5.53/(T_B(^{12}\mathrm{CO}) + 0.819)]}.
\end{equation}
where $T_B(^{12}\mathrm{CO})$ is the brightness temperature of $^{12}$CO (J = 1-0) at 
the peak of C$^{18}$O (J = 1-0).
The excitation temperature of SFO 22 is found to be 27.1$\pm$1.8 K. 
Fig. \ref{pv map sfo22} shows the position-velocity diagrams of $^{12}$CO (J = 1-0), 
$^{13}$CO (J = 1-0), and C$^{18}$O (J = 1-0) along declination through the peak position
of C$^{18}$O (J = 1-0). The line widths of  $^{12}$CO (J = 1-0), $^{13}$CO (J = 1-0), and 
C$^{18}$O (J = 1-0) are roughly $\sim$ 2.0 km $\mathrm{s^{-1}}$, $\sim$ 1.5 km $\mathrm{s^{-1}}$, 
and $\sim$ 1.0 km $\mathrm{s^{-1}}$, respectively. 
Line widths of SFO 22 are relatively narrow compared to other BRCs.
In the millimeter and sub-millimeter molecular line survey of BRCs by \cite{2002ApJ...577..798D}, many BRCs have CO line widths of $\gtrsim$ 5 km $\mathrm{s^{-1}}$. These large line widths of BRCs may be attributed to large velocity dispersion due to turbulence and outflow activities. The observed narrow line widths of SFO 22 could suggest that the influence from the photoionization-induced shocks may have already disappeared, and SFO 22 has now reached the phase of quasi-stationary equilibrium predicted in the radiation implosion model. 

\begin{figure}
 \begin{center} 

 \plotone{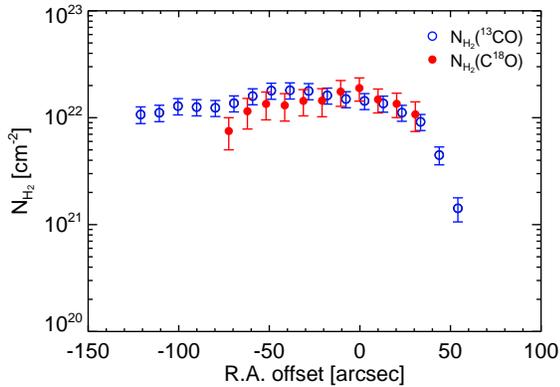}

\caption{Column density profiles of SFO 22 along the declination. 
      The open circles (colored blue in the online version) and filled circles
      (colored red in the online version) label the column densities derived from $^{13}$CO (J = 1-0) emission and C$^{18}$O (J = 1-0) emission, respectively. 
      \newline 
     (A color version of this figure is available in the online journal.)}
\label{column density sfo22}
\end{center}
\end{figure}

Figure \ref{column density sfo22} shows column density profiles of SFO 22 along declination, which is 
almost parallel to the direction to the exciting star.  
We derive column density distribution of the cloud using data from this observation.
Assuming that the excitation temperatures of the $^{13}$CO and C$^{18}$O lines are the same 
as the $^{12}$CO line, we can calculate optical depth of $^{13}$CO and C$^{18}$O using
\begin{equation}
 \tau_{13} = - \ln \left( 1 - \frac{T_B(\mathrm{^{13}CO})}
                    {5.29 (1/(\exp(5.29/T_{ex}) - 1)) - 0.164)}  \right)
\end{equation}
and
\begin{equation}
 \tau_{18} = - \ln \left( 1 - \frac{T_B(\mathrm{C^{18}O})}
                    {5.27 (1/(\exp(5.27/T_{ex}) - 1)) - 0.166)}  \right),
\end{equation}
respectively.
The column densities of $^{13}$CO and C$^{18}$O molecules can be derived as 
\begin{equation}
 N(\mathrm{^{13}CO}) = 2.42 \times 10^{14} 
               \frac{\tau_{13} \Delta v_{13} T_{ex}}{1-\exp[-5.29/T_{ex}]} 
\end{equation}
and
\begin{equation}
 N(\mathrm{C^{18}O}) = 2.42 \times 10^{14} 
               \frac{\tau_{18} \Delta v_{18} T_{ex}}{1-\exp[-5.27/T_{ex}]},
\end{equation}
where $\Delta v_{13}$ and $\Delta v_{18}$ are line widths of the $^{13}$CO (J = 1-0) and 
C$^{18}$O (J = 1-0) emission, respectively. 
The column density of $^{13}$CO is converted to the column density of $\mathrm{H_2}$ 
by assuming an abundance ratio of $N({\rm H_2})/N({\rm ^{13}CO}) = 5.0 \times 10^5$ \citep{1978ApJS...37..407D}. 
Under the environment where molecular clouds are illuminated by strong UV radiation, abundance ratio of 
$N({\rm H_2})/N({\rm C^{18}O})$ are thought to be affected by selective destruction by UV radiation \citep{1985ApJ...290..615G}.
We follow \cite{2009A&A...500.1119N} to obtain the column density of $\mathrm{H_2}$ from the column density of C$^{18}$O.
The column densities of $^{13}$CO and C$^{18}$O can be formulated by least-square fitting as
\begin{equation}
 N(\mathrm{C^{18}O}) = 3.91 \times 10^{-2}  \times N(\mathrm{^{13}CO}).
\end{equation}
 An abundance ratio of $N({\rm H_2})/N({\rm C^{18}O}) = 1.28 \times 10^7$ can be derived.
This value is slightly larger than the standard value of $6.0 \times 10^6$ for molecular clouds not associated with \ion{H}{2} regions \citep{1982ApJ...262..590F}.
Figure \ref{column density sfo22} shows that the column density profiles obtained here are nearly flat with the column density of $\sim 10^{22}$ $\rm cm^{-2}$.
Adopting a distance of 460 pc to SFO22, the total masses traced by $^{13}$CO emission and C$^{18}$O 
emission are $M_{13} = 12.0$ $\mathrm{M_{\sun}}$ and $M_{18} = 5.3$ $\mathrm{M_{\sun}}$, respectively.

\section{Model Description}
\label{sect model}

Our numerical model is based on the analytic model of \citet{1990ApJ...354..529B} for a photo-evaporating cloud in the quasi-stationary equilibrium state with a polytropic equation of state. The polytropic gas is a good approximation for the cases where either gas pressure or magnetic pressure is dominant. In an actual bright-rimed cloud, these two quantities may have been comparable before the compression by radiation-driven implosion alter the state, and FUV radiation will affect the thermal structure through photoelectric heating and photodissociation of molecular coolants. Here we adopt a more realistic model with the inclusion of thermal pressure and magnetic pressure explicitly and the heating due to FUV radiation.  

\subsection{Density Structure}
\begin{figure}
  \begin{center}
      \epsscale{0.95}
    \plotone{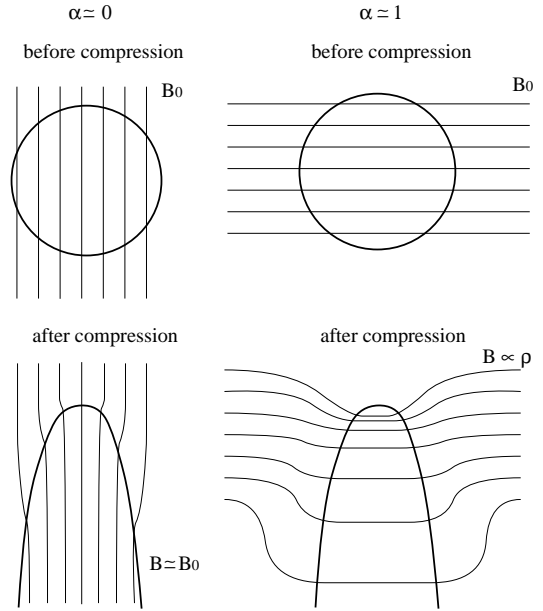}
  \end{center}
       \caption{Schematic figures of compression of magnetized cloud due to radiation-driven implosion for two extreme cases. 
           UV radiation propagates downward, and the cloud is compressed only along this direction in both cases. 
           (left) The magnetic field is parallel to the direction of UV radiation.  
           (right) The magnetic field is perpendicular to the direction of UV radiation. } 
     \label{figure alpha}
\end{figure}

Total pressure of the gas is expressed as 
\begin{eqnarray}
 P_{tot} &=& P_{th} + P_{mag} \nonumber \\
   &=& \frac{c_s^2 \rho}{\gamma} + \frac{B^2}{8 \pi},
\label{total pressure}
\end{eqnarray}
where, $P_{th}$, $P_{mag}$, $c_s$, $\rho$, $\gamma$, and $B$ are the thermal pressure, the magnetic pressure, 
the sound speed, the density of the cloud, the ratio of specific heats, and the magnetic field strength, respectively.
A $\gamma = 5/3$ is adopted, which is appropriate for the molecular clouds because the molecular hydrogen behaves like 
a monoatomic gas at temperature $\lesssim$ 100 K. We assume ideal gas, so that the sound speed $c_s$ is related with the gas temperature T as 
\begin{eqnarray}
    c_s = \sqrt{\frac{\gamma k_B T}{\mu m_{\rm H}}},
\end{eqnarray}
where $k_B$ and $\mu$ are the Boltzmann constant and mean mass per nucleus in unit of the hydrogen mass 
$m_{\rm H} = 1.67 \times 10^{-24}$ g, respectively, and $\mu = 1.15$.
For simplicity, the effects of magnetic field is approximated through 
\begin{equation}
 B = B_0 \left( \frac{\rho}{\rho_0} \right)^\alpha,
  \label{dependence of magnetic field}
\end{equation}
where $B_0$ and $\rho_0$ are magnetic field strength and density of the cloud before undergoing compression by radiation-driven implosion. 
The exponent $\alpha$, which ranges from 0 to 1, is a parameter representing how much magnetic field is trapped in the gas during the compression. 
Figure \ref{figure alpha} shows schematic drawings of two extreme cases by assuming that compression of a cloud perpendicular to the direction of radiation is small. 
If the cloud is compressed along the magnetic field, the magnetic field strength will hardly change during compression, i.e. $\alpha \simeq 0$. 
On the other hand, if the cloud is compressed perpendicular to the magnetic field, 
the strength of the field increases as $B \propto \rho$ during compression, i.e. $\alpha \simeq 1$. The value of $\alpha$ depends on the initial configuration 
of magnetic field and the shape of cloud.
We leave $\alpha$ an open parameter as it is hard to determine an accurate value without launching MHD simulations.
We also neglect the diffusion of magnetic field due to the longer timescale of ambipolar diffusion compared to the dynamical timescale of the radiation-driven implosion.

\begin{figure}
  \begin{center}
      \epsscale{1.0}
    \plotone{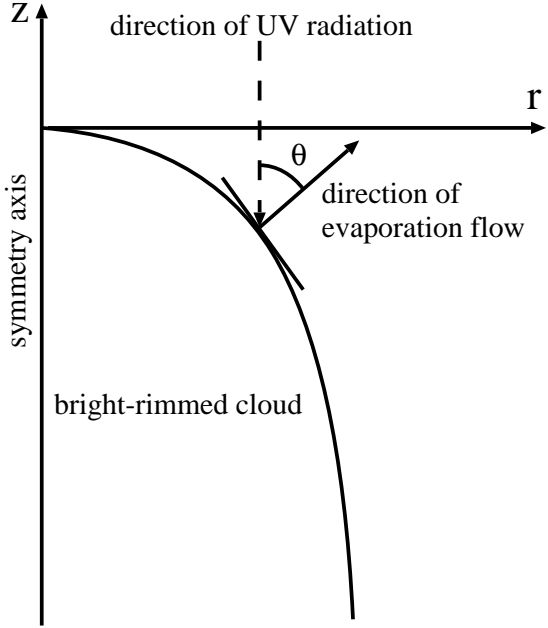}
  \end{center}
       \caption{The coordinate system used in our numerical models. The exciting star is assumed far above. 
Ionizing UV radiation is parallel to the symmetric axis. The flow of evaporation streams along the surface normal. 
 The angle between surface normal and UV radiation is denoted by $\theta$.}
     \label{schematic images}
\end{figure}

We calculate structure of the cloud in axisymmetry along the line to the exciting star.
We also assumed that distance from the cloud to the exciting star is larger than size of the cloud, so that UV radiation field can be treated as planar. 
Figure \ref{schematic images} illustrates the coordinates system we use in this paper. 
UV radiation propagates downward parallel to the z-axis, and the ionized gas evaporates off the cloud surface 
with angle $\theta$ to z-axis. \citet{1990ApJ...354..529B} showed that the position of the cloud surface can be approximated as  
\begin{equation}
  z_{sur} = R_c a^{-2} \ln \cos(a r/ R_c),
  \label{eq cloud surfac}
\end{equation}
where $R_c$ and $a$ are the curvature radius at $z=0$ and the cloud width parameter defined in 
\citet{1990ApJ...354..529B}, respectively. 
The method used to determine $a$ and $R_c$ is described in section \ref{sect iteration}.

In this study, we assume that all part of the cloud is subject to the same acceleration through the rocket effect of 
the photo-evaporation flow, as the cloud is in quasi-stationary equilibrium state.
The equation of hydrostatic equilibrium for the cloud is
\begin{equation}
  \frac{d P_{tot}}{d z} = g \rho,
  \label{pressure balance}
\end{equation}
where $g$ is the acceleration of the cloud.
With the  equation (\ref{total pressure}) and (\ref{dependence of magnetic field}), this equation 
can be rewritten as
\begin{equation}
  \frac{d \rho}{d z} = \frac{g \rho - 2 c_s \rho / \gamma (d c_s / d z)}{c_s^2 / \gamma + 2 \alpha B_0^2 \left( \rho \right / \rho_0)^{2 \alpha - 1}/ 8 \pi \rho_0} .
  \label{derivative of the density}
\end{equation}
The density distribution inside the cloud is determined by solving this equation with appropriate boundary conditions. 

The boundary conditions at cloud surface are determined using jump conditions for D-critical ionization front. 
Therefore, pressure at cloud surface is given by
\begin{equation}
    P_{sur}(r) = 2 \mu m_{\rm H} c_i  F_{UV}(r) \cos{\theta},
 \label{surface pressure}
\end{equation}
where $c_i = 10 \, {\rm km \, s^{-1}}$ and $F_{UV}(r)$ are the sound speed of ionized gas and
ionization photon flux reaching ionization front, respectively, and the width of the ionization front is negligible.
Since some part of the incident ionizing UV photons is consumed by recombined hydrogens in the photoevaporation flow, 
ionizing UV photon flux arriving at the cloud surface is written as
\begin{equation}
   F_{UV}(r) = F_i - \int_{z_{sur}}^{\infty} \alpha_{B} n_e(r,z) n_p(r,z) \, dz, 
   \label{uv flux at surface}
\end{equation} 
where $F_i$, $\alpha_B=2.7 \times 10^{-13}$ ${\rm cm^3 \, s^{-1}}$, $n_e$, and $n_p$ are the incident ionizing photon flux, 
the hydrogen electronic recombination coefficient into the excited state, the electron number density, and the proton number density, 
respectively. 
Following \citet{1990ApJ...354..529B}, we introduce a dimensionless parameter $\omega$ that represents the effective fractional 
thickness of the recombination layer. 
Equation (\ref{uv flux at surface}) can be rewritten as 
\begin{equation}
   F_{UV}(r) = F_i - \omega(r) \alpha_B n_{II}^2(r=0) R_c,
\end{equation}
where $n_{II}$ is the hydrogen number density just behind the ionization front.
If we assume the ionized gas streams away from the ionization front with the velocity of sound speed, and the
stream line is normal to the cloud surface, we can derive the form,
\begin{eqnarray}
    \omega(r)   & \simeq & 3 \omega(0) \int_0^r \left[ \frac{r' P_{tot}(r')}{r P_{tot}(0)} \right]^2 \times  \nonumber \\
                & &  \frac{R(r')}{[R(r') \sin{\theta'} + r -r'] \cos{\theta'}} \frac{dr'}{R_c},
\end{eqnarray}
where $R(r')$ is the curvature radius of the cloud surface at $r=r'$. 
We adopt $\omega$ for a spherical cloud with radius $R_c$ at the cloud tip:
\begin{equation}
   \omega(0)  = \frac{q (q - 1)}{\psi},
\end{equation}
where $q$ and $\psi$ are the ratio of incident ionizing photon flux $F_i$ to ionizing photon 
flux reaching ionization front $F_{UV}$ and the photoevaporation parameter, respectively. 
The photoevaporation parameter is defined as 
\begin{equation}
  \psi = \frac{\alpha_B F_i R_c}{c_i^2}.
\end{equation}
The value of $\psi$ in our numerical models ranges from 74 to 104 depending on the model parameters.
\citet{1978ppim.book.....S} derived an analytic estimation of $q$ as
\begin{equation}
   q = \frac{1 + \left( 1 + 1.5 \psi^{1/4}\right)^2 }{8}.
\end{equation}

The density and pressure distributions of the cloud are characterized by scale height defined as $h_c = P_{tot}(z=0) 
/ g \rho(z=0)$. If $\psi \gg 1$, which is the case for SFO 22, $h_c$ is related to $R_c$ as $R_c = 0.5 h_c$.  

\subsection{Thermal and Chemical Model}

We solve the reaction networks for the species of H$_2$, CO, C$^+$, O, and the electron $e$.
We adopt the simplified reaction model as described in \citet{1997ApJ...482..796N} to determine the abundance of CO 
molecules.  In this reaction model, C$^+$ is directly converted to CO without accounting explicitly for the 
intermediate reactions. The formation rate of CO molecules is expressed as
\begin{equation}
    R_{\mathrm{CO}} = 5 \times 10^{-16} n(\mathrm{C^+}) n(\mathrm{H_2}) \beta,
\end{equation}
where $n(\mathrm{C^+})$ and $n(\mathrm{H_2})$ are the number densities of C$^+$ and H$_2$, respectively.
The coefficient $\beta$ is defined as
\begin{equation}
 \beta = \frac{5 \times 10^{-10}X(\mathrm{O})}{5 \times 10^{-10} X(\mathrm{O}) + D_{\mathrm{CH_x}} / n(\mathrm{H_2})},
\end{equation}
where $X(\mathrm{O})$ and $D_{\mathrm{CH_x}}$ are the fractional abundance of oxygen and the total photodissociation 
rate of both CH and CH$_2$. This total photodissociation rate is written as
\begin{equation}
  D_{\mathrm{CH_x}} = 5 \times 10^{-10} G_0 \exp (-\tau_{UV}), 
\end{equation}
where $G_0$ and $\tau_{UV}$ are the intensity of the incident FUV radiation in terms of the Habing 
interstellar radiation field \citep{1968BAN....19..421H} and the optical depth in UV range, respectively.
The optical depth $\tau_{UV}$ is related to the visual extinction $A_v$ by $\tau_{UV}=2.5 A_v$.
We adopt the conversion factor between the visual extinction and the total hydrogen column density as
$X_{A_V} = A_V / N_{\rm H} = 6.3 \times 10^{-22}$ mag cm$^2$ in this paper. 
The photodissociation rate of CO by the FUV radiation is written as  
\begin{equation}
    D_{\mathrm{CO}} = 10^{-10} G_0 \exp(-\tau_{UV}) n(\mathrm{CO}),
\end{equation}
where $n(\mathrm{CO})$ is the number density of CO.
The abundance of CO is calculated by solving the equation of formation and dissociation balance  
\begin{equation}
    R_{\mathrm{CO}} -  D_{\mathrm{CO}}  = 0.
\label{reaction of CO}
\end{equation}

Abundances of other species are calculated as follows.
The cloud is assumed to be composed of molecular hydrogen, so that the number density of H$_2$ is written as 
\begin{equation}
    n(\mathrm{H_2}) = 0.5 n, 
\end{equation}
where $n$ is the number density of hydrogen nuclei and related with mass density as
\begin{equation}
    n = \frac{\rho}{\mu m_{\rm H}} .
\label{eq number density}
\end{equation}
We assume that carbon exists in ionized form in the cloud as carbon is easily photoionized by FUV radiation owing to its lower 
ionization energy (11.2 eV) compared to that of hydrogen. 
Although this assumption may cause overestimation of cooling rate through C$^+$ in the deep inner region of cloud where 
FUV is strongly attenuated and rotational line emission from CO is a more important cooling process. Our assumption does not affect 
the numerical results, however. 
The number density of C$^+$ is calculated by
\begin{equation}
    n(\mathrm{C^+}) = X(\mathrm{C_{tot}}) n -  n(\mathrm{CO}). 
\end{equation}
where the elemental abundance of carbon is taken to be $X(\mathrm{C_{tot}}) = 10^{-4}$.
We assume that oxygen exists in atomic form in the cloud, because its ionization energy (13.6 eV) is similar to 
that of hydrogen.
The number density of oxygen is calculated by
\begin{equation}
    n(\mathrm{O}) = X(\mathrm{O_{tot}}) n -  n(\mathrm{CO}). 
\end{equation}
where the elemental abundance of oxygen is taken to be $X(\mathrm{O_{tot}}) = 2.0 \times 10^{-4}$.
Constant electron fraction $n_e / n = 10^{-7}$ is also assumed to calculate number density of electron.

For the heating processes in the cloud, we consider photoelectric heating and cosmic ray heating. 
The photoelectric heating rate is \citep{1994ApJ...427..822B}
\begin{equation}
  \Gamma_{pe} = 10^{-24} \epsilon \, G_0 \exp(-\tau_{UV}) n ~ \mathrm{ergs~ cm^{-3}~ s^{-1}},
  \label{eq heat pe}
\end{equation}
where $\epsilon$ is the photoelectric heating efficiency
\begin{eqnarray}
  \epsilon &=& \frac{4.87 \times 10^{-2}}{1 + 4 \times 10^{-3} (G_0 \exp(-\tau_{UV}) T^{1/2} / n_e)^{0.73}} \nonumber \\
            && + \frac{3.65 \times 10^{-2} (T/10^4)^{0.7}}{1 + 2 \times 10^{-4} (G_0 \exp(-\tau_{UV}) T^{1/2} / n_e)}.
  \label{eq efficiency pe}
\end{eqnarray}
Cosmic ray heating become the dominant heating process in the inner region where FUV radiation does not penetrate.
Cosmic ray heating rate is given by 
\begin{equation}
  \Gamma_{cr} = \zeta_p(\mathrm{H_2}) \Delta Q_{cr}  n(\mathrm{H_2})~ \mathrm{ergs~ cm^{-3}~ s^{-1}},
  \label{eq heat cr}
\end{equation}
where $\zeta_p(\mathrm{H_2})$ is the primary cosmic ray ionization rate of $\mathrm{H_2}$ and 
$\Delta Q_{cr}$ is the energy deposited as heat as a result of this ionization. 
We adopt values of $\zeta_p(\mathrm{H_2}) = 7.0 \times 10^{-17}$ $\mathrm{s^{-1}}$ \citep{1986ApJS...62..109V} 
and $\Delta Q_{cr} = 20$ eV \citep{1978ApJ...222..881G}.  

For the cooling processes in the cloud, we consider radiation from the CO rotational transitions, collisionally excited line 
emission from C${^+}$ and O, and collisional heat transfer between gas and dust.
The cooling rate due to CO, $\Lambda_{\mathrm{CO}}$ is taken from the tabulated cooling function computed 
by \citet{1995ApJS..100..132N} for $T \leq $ 100 K and \citet{1993ApJ...418..263N} for $T>$ 100 K.
The cooling rates due to collisional excitation of C$^{+}$ and O are taken from \citet{1997ApJ...482..796N}. 
The cooling rate due to C$^{+}$ is 
\begin{eqnarray}
  \Lambda_{\mathrm{C^+}} = [1.1 \times 10^{-23} n(\mathrm{C^+}) n 
                              \exp(-92/T)]   \nonumber \\
                              /  \{1 + (n(\mathrm{H_2})/n_{crit}) [1 + 2 \exp(-92/T)]\},
\end{eqnarray}
where the critical number density is taken to be $n_{crit} = 3 \times 10^3~ \mathrm{cm^{-3}}$.
The cooling rate due to O is 
\begin{eqnarray}
  \Lambda_{\mathrm{O}} = 5.0 \times 10^{-27} n(\mathrm{O}) n 
                              [24 \exp(-228/T)]   \nonumber \\
                          + 7 \exp(-326/T)] T^{1/2}.
\end{eqnarray}
The cooling rate by the dust grain is \citep{1989ApJ...342..306H}
\begin{eqnarray}
    \Lambda_{dust} = [1.2 \times 10^{-31} n^2 \left( \frac{T}{1000 \, \mathrm{K}} \right)^{1/2} \left( \frac{100 \mathrm{\AA}}{a_{min}} \right)^{1/2} 
   \nonumber \\ 
            \times   [1 - 0.8 \exp(-75/T)] (T - T_{dust}),
\end{eqnarray}
where $a_{min}$ and $T_{dust}$ are the minimum radius of grains and the dust temperature, respectively. 
We used $a_{min} =$ 100 $\mathrm{\AA}$ to calculate the cooling rate by dust grains.  
The dust temperature is calculated following the method by \cite{Hollenbach1991}.

\subsection{An Iterative Procedure for Obtaining Numerical Solution}\label{sect iteration}

An iterative procedure is used to achieve the density distribution of the cloud. 
We use a uniform grid of 200 (radial) $\times$ 1000 (axial) cells.
The grid spacing is $6.5 \times 10^{-4}$ pc.
For presentation, we symmetrize the results obtained on a grid of 400 $\times$ 1000 cells. 
In order to fit the width of the cloud with that of SFO 22, the cloud width parameter $a$ is calculated by substituting 
$r = 0.13$ pc and $z = - 0.6$ pc into equation (\ref{eq cloud surfac}).
The procedure is as follows.
(i) Equation (\ref{derivative of the density}) is numerically solved by Runge-Kutta method 
using current temperature and sound speed. (ii) Optical depth to the cloud surface is calculated. 
(iii) Chemical reaction network is solved to determine abundances of included species. 
(iv) The temperature and sound speed are updated using new chemical abundances.
The thermal equilibrium is assumed to determine the temperature of cloud:
\begin{equation}
  \Gamma_{cr} +  \Gamma_{pe} - \Lambda_{\mathrm{CO}} - \Lambda_{\mathrm{C^+}} - \Lambda_{\mathrm{O}}  - \Lambda_{dust}= 0.
\end{equation}
These processes are repeated until the converged solution is achieved. We have confirmed that relative errors between the two consecutive steps are smaller than $10^{-2}$ for all models in this paper.

Once the curvature radius of the cloud $R_c$ is given, we can determine the position of cloud surface from equation 
(\ref{eq cloud surfac}) and calculate structure of the cloud using iterative procedure described above. 
Since we assume that the cloud is in an equilibrium state, we adopt $R_c$ which minimizes the pressure gradient along r-direction. 
We change $R_c$ from 0.02 pc to 0.11 pc in increments of 0.001 pc, and calculate structure of the cloud for each $R_c$.
Then, we integrate deviation of total pressure $P_{tot}(r,z)$ from the average total pressure over all cells at 
the same z coordinates $P_{ave}(z)$,
\begin{equation}
  \frac{\int \left| \left(P_{tot}(r, z) - P_{ave}(z) \right) / P_{ave}(z) \right| dV}{\int dV}.
\end{equation}
We adopt the structure of cloud for which this value is minimized.


\section{Results}\label{sect results}
\begin{figure}
    \begin{center}
    \epsscale{0.95}
    \plotone{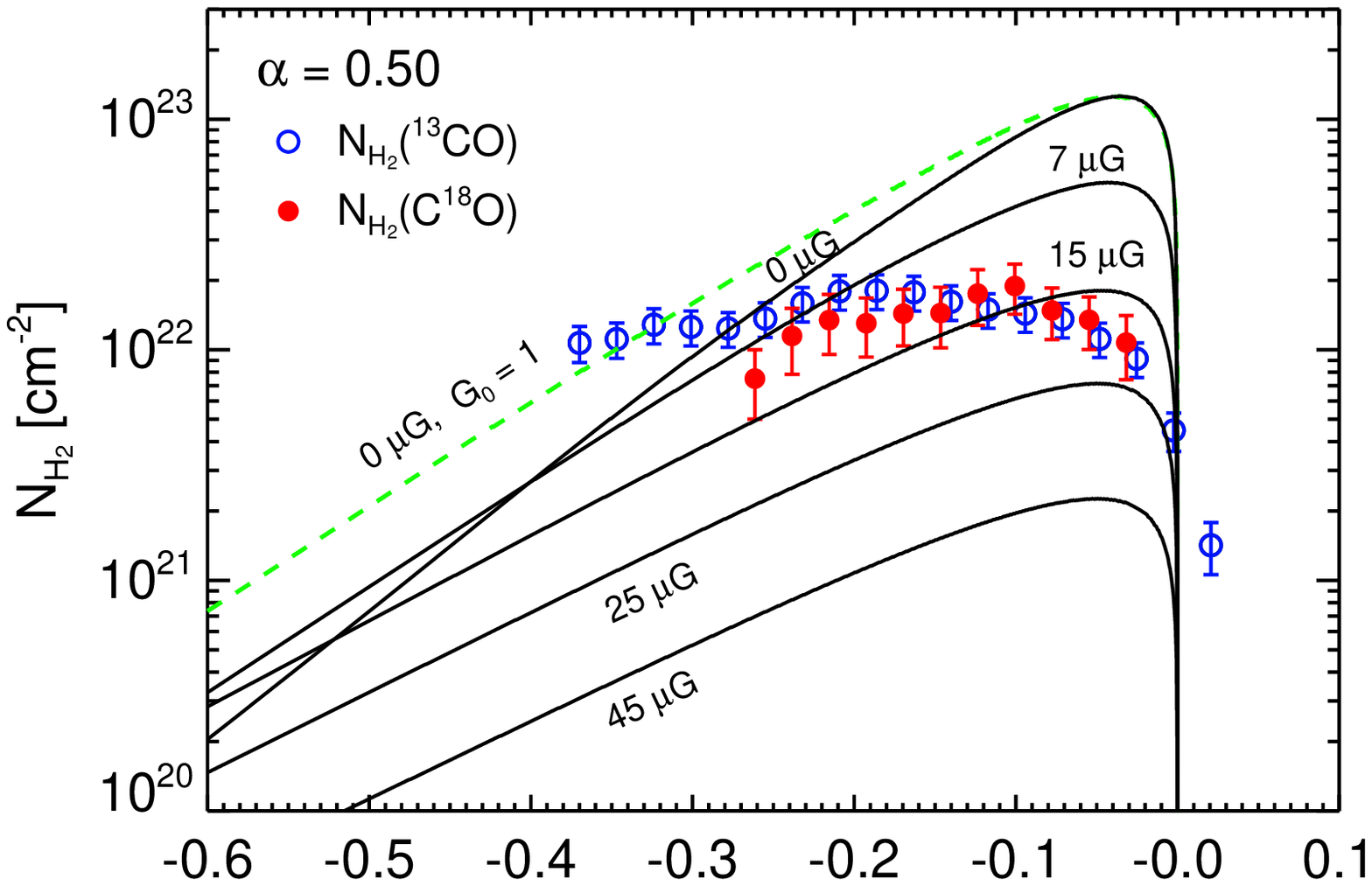}
    \plotone{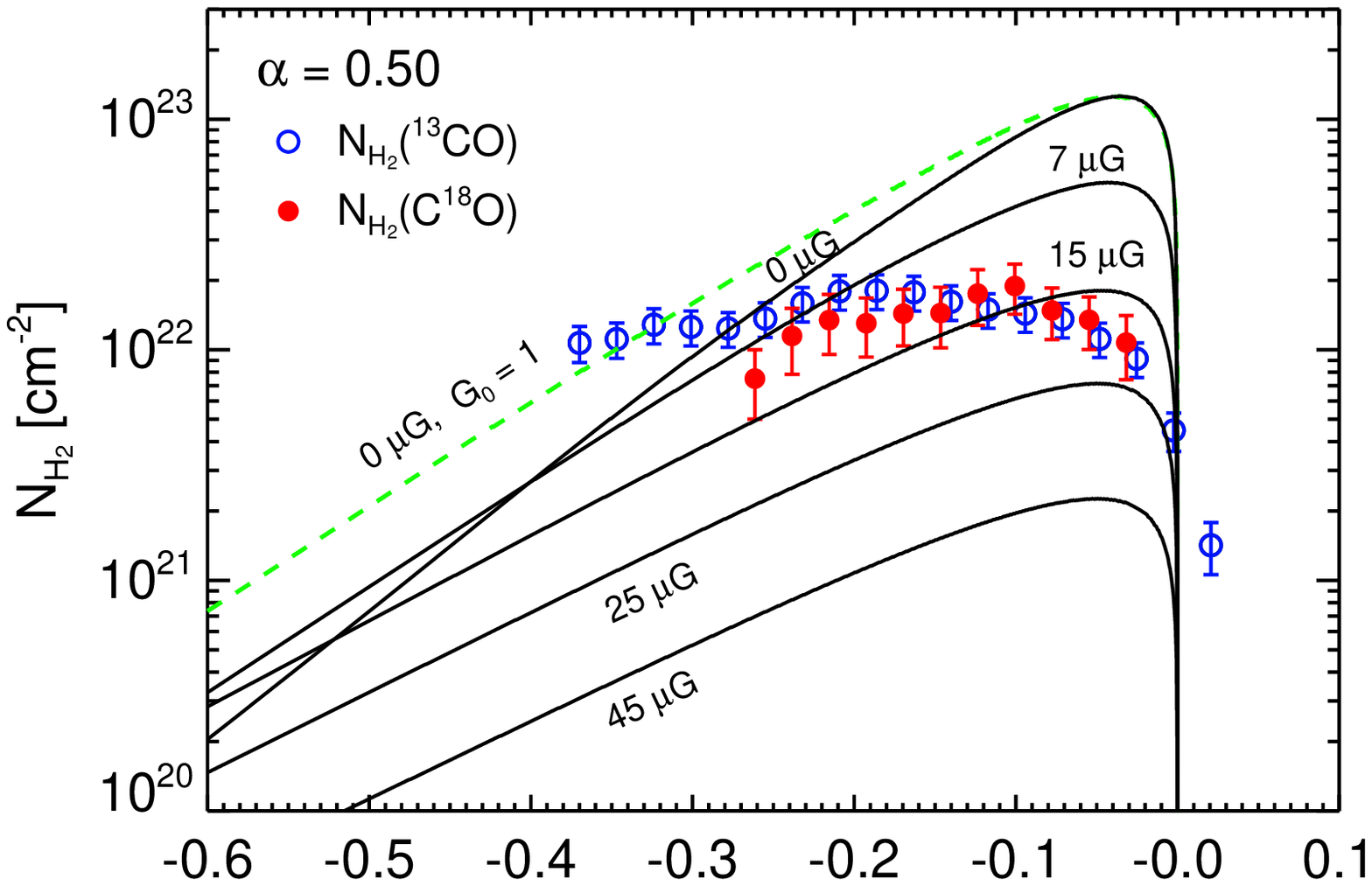}
    \plotone{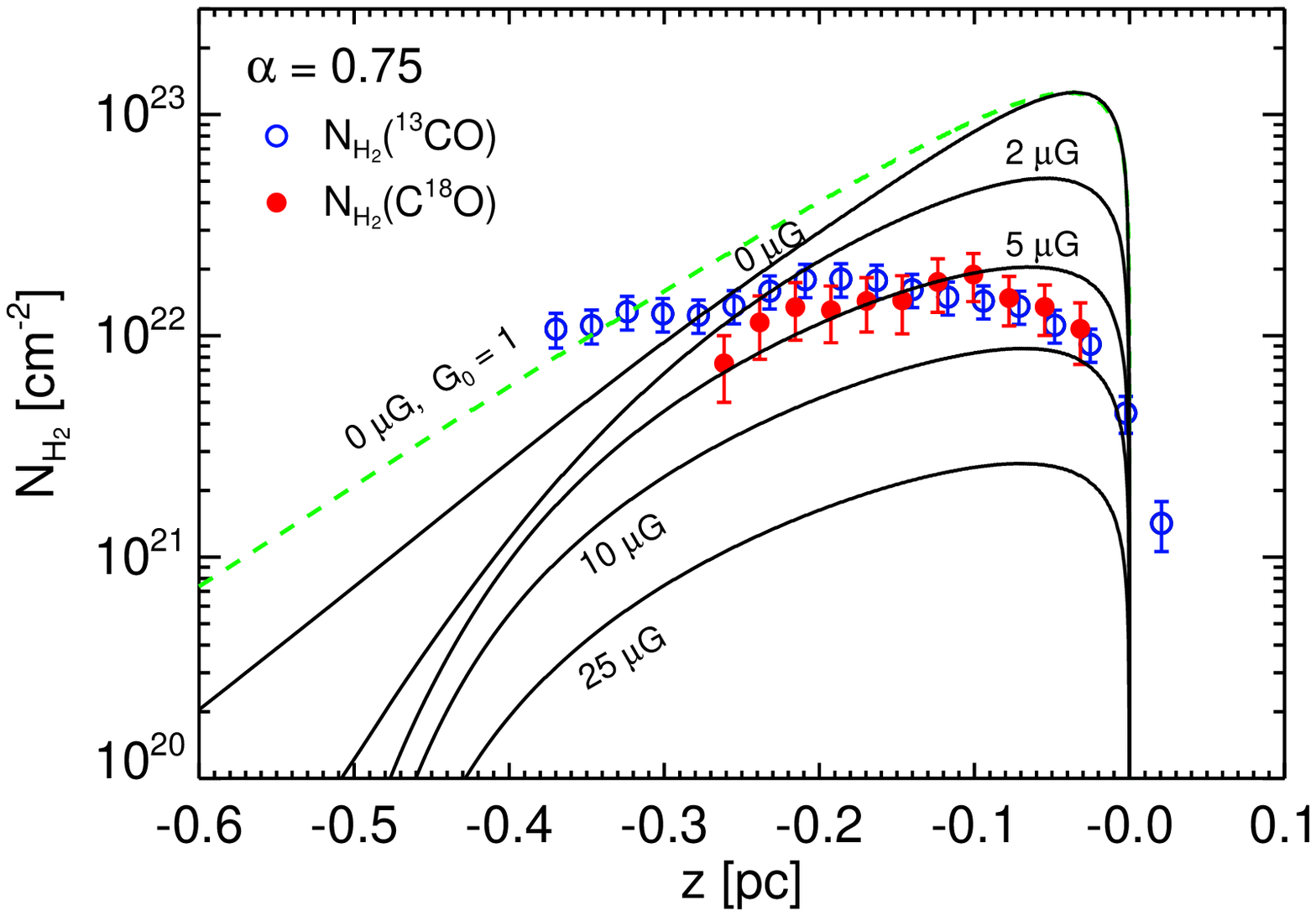}
    \end{center}
       \caption{Comparisons of the column density profiles along the z-axis with the column densities derived from observations. 
      The top, middle, and bottom panels show the results with $\alpha=0.25$,
      0.50, and 0.75, respectively. The dashed line (colored green in the
      online version) represents the column density profile of model A. Column
      densities of SFO 22 are plotted with same symbols as Figure \ref{column density sfo22}. \newline 
      (A color version of this figure is available in the online journal.)}
     \label{fig column density profiles}
\end{figure}

We present results of our numerical models in this section.
Model parameters of the run  are summarized in Table \ref{tbl parameters}. 
To compare with our observational results, intensities of the incident ionizing UV and FUV radiation 
were determined to correspond to those expected in the region where SFO 22 is located.
Since the spectral type of the primary exciting star of SFO 22 is O7V, 
we adopted the values of $\log S_{UV} =  48.76$ $\mathrm{s^{-1}}$ and $\log S_{FUV} =  48.76$ $\mathrm{s^{-1}}$ 
as the UV and FUV photon luminosities \citep{1996ApJ...460..914V, 1998ApJ...501..192D}.   
These luminosities and the distance from the exciting star to SFO 22 of 6.5 pc give the incident ionizing UV flux 
of $F_{i} = 1.192 \times 10^9$ $\mathrm{cm^{-2}\, s^{-1}}$ and incident FUV flux of $G_0 = 94$ in terms of  
the Habing field $F_H = 1.21 \times 10^7$ $\mathrm{cm^{-2}\, s^{-1}}$ \citep{1996ApJ...458..222B}. 
We used these values in all models except model A. The number density of the cloud before undergoing compression by 
radiation-driven implosion was assumed to be $n_0 = \rho_0 / \mu m_{\rm H} = 10^3~ \mathrm{cm^{-3}}$ in all models.
Some physical values obtained from our results are summarized in Table \ref{tbl results}.
In Figure \ref{fig column density profiles}, column density profiles along the z-axis for all models  
are plotted with column densities of SFO 22 derived from our observations.

\begin{table}
\begin{center}
\caption{Model parameters\label{tbl parameters}}
\begin{tabular}{ccccc}
\tableline
\tableline
Model &  $\alpha$ & $B_0$ [$\mu$G] &  $G_0$ & $F_{i}$ [$\mathrm{cm^{-2}\, s^{-1}}$] \\
\tableline

A   & -    &  0 &   1.0  & $1.192 \times 10^{9}$  \\
B   & -    &  0 &  94.0  & $1.192 \times 10^{9}$   \\
C1  & 0.25 & 25 &  94.0  & $1.192 \times 10^{9}$    \\
C2  & 0.25 & 45 &  94.0  & $1.192 \times 10^{9}$    \\
C3  & 0.25 & 60 &  94.0  & $1.192 \times 10^{9}$    \\
C4  & 0.25 & 80 &  94.0  & $1.192 \times 10^{9}$    \\
D1  & 0.50 &  7 &  94.0  & $1.192 \times 10^{9}$    \\
D2  & 0.50 & 15 &  94.0  & $1.192 \times 10^{9}$    \\
D3  & 0.50 & 25 &  94.0  & $1.192 \times 10^{9}$    \\
D4  & 0.50 & 45 &  94.0  & $1.192 \times 10^{9}$    \\
E1  & 0.75 &  2 &  94.0  & $1.192 \times 10^{9}$    \\
E2  & 0.75 &  5 &  94.0  & $1.192 \times 10^{9}$    \\
E3  & 0.75 & 10 &  94.0  & $1.192 \times 10^{9}$    \\
E4  & 0.75 & 25 &  94.0  & $1.192 \times 10^{9}$   \\
\tableline
\end{tabular}
\end{center}
\end{table}

\subsection{Effects of Strong FUV Radiation and Magnetic Field}\label{sect FUV and B}
\begin{figure*}
  \begin{center} 
  
  \epsscale{0.9}
  \plotone{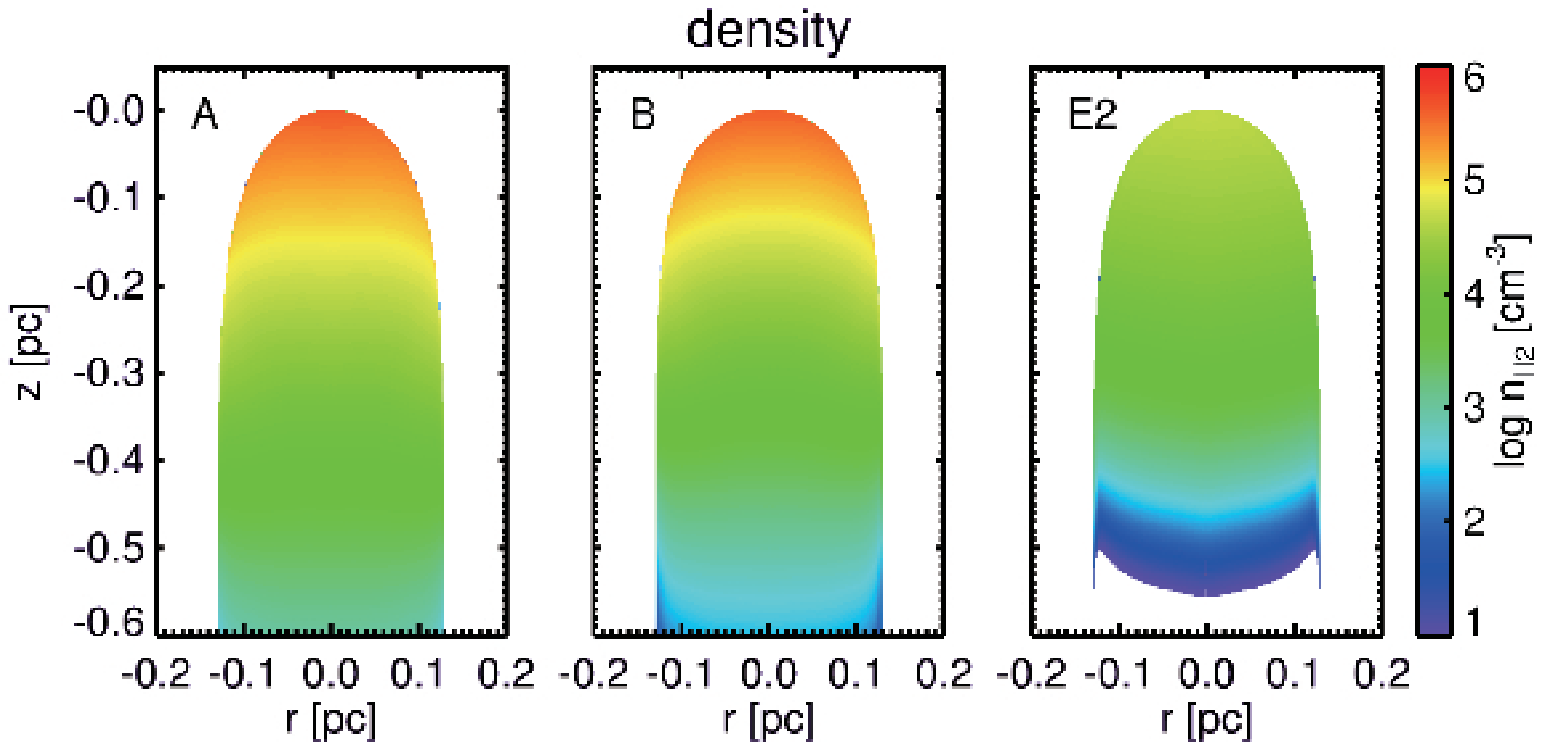}\\
  \plotone{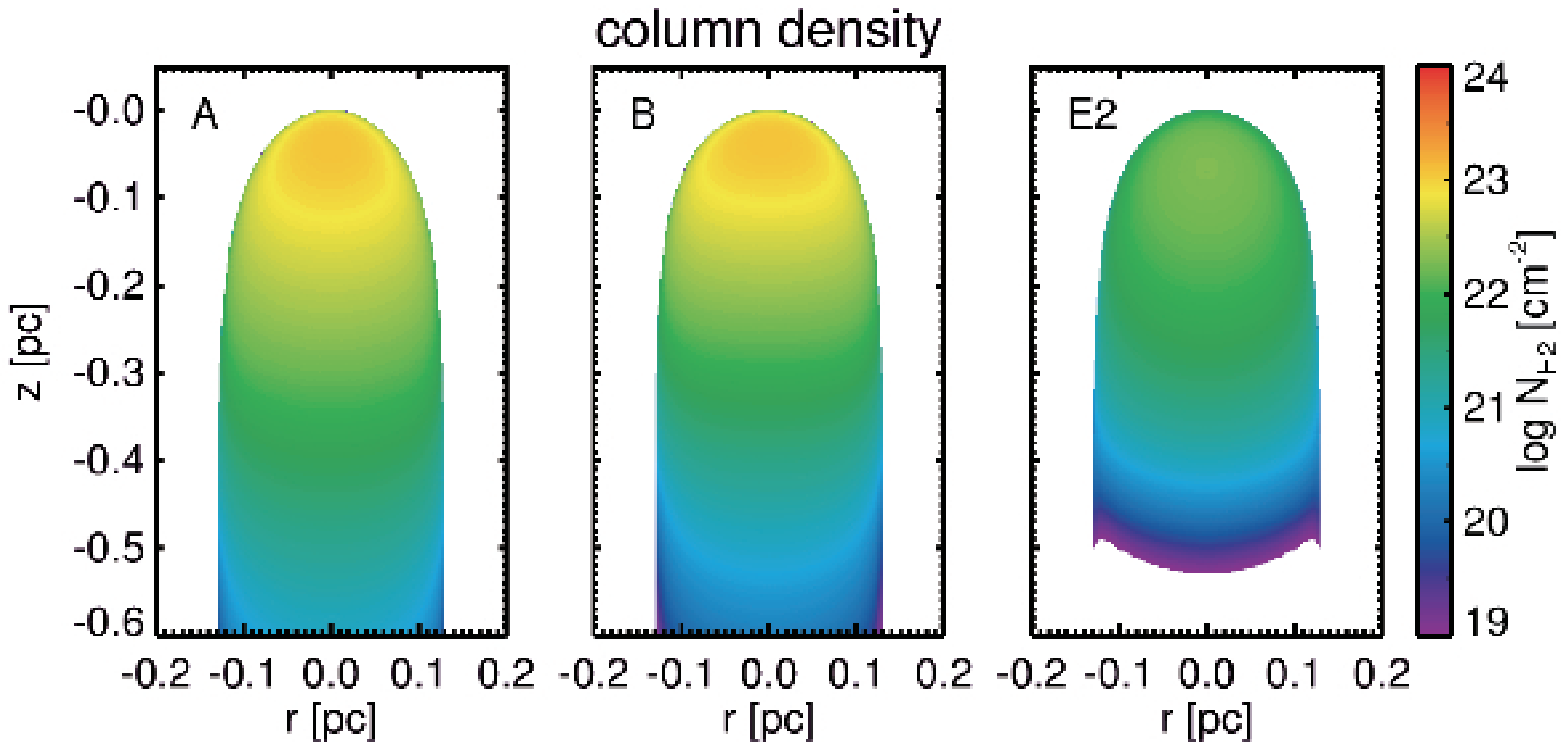}\\
  \plotone{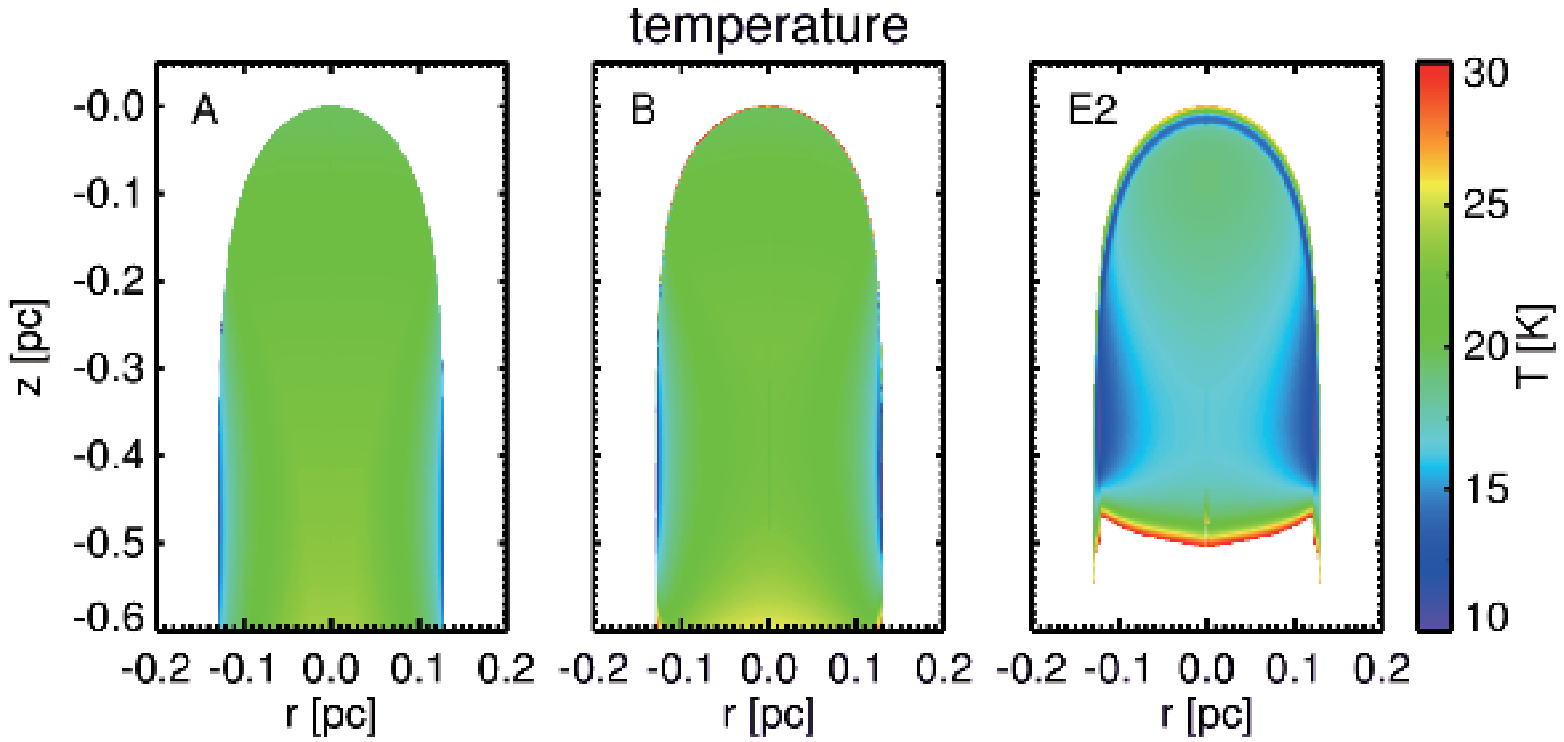}
  
  \end{center}
  \caption{The density, the column density, and the temperature distributions in model A (left), B (center), and, E2 (right).}
  \label{fig comparison three models}
 \end{figure*}

In this subsection, we describe results of following three typical models, A, B, and E2 to show how strong FUV radiation 
and magnetic field affect structures of clouds. 
Figure \ref{fig comparison three models} shows the densities, the column densities, and the temperatures of these models.
In model A, as a reference, we calculated the structure of the cloud without magnetic field, assuming strength of average 
interstellar FUV radiation field of $G_0 = 1$.
In model B, to see the effects of strong FUV radiation, we calculated the structure of the cloud assuming strength of 
FUV radiation expected at the region where SFO 22 is located, but the effects
of magnetic field was not included.
In model E2, we included not only the effects of strong FUV radiation but also the effects of magnetic field.  
The initial magnetic field strength $B_0$ and $\alpha$ were set to be 5 $\mu$G and 0.75, respectively. 

Figure \ref{fig column density profiles} shows that the column density distribution 
of model A is much different from that of observed cloud.
The peak column density of $1.26 \times 10^{23}$ $\mathrm{cm^{-2}}$ is one
order of magnitude higher than that of 
observed cloud ($ \sim 2 \times 10^{22}$ $\mathrm{cm^{-2}}$). In addition, slope of column density profile 
is steeper than observations.
Figure \ref{fig comparison three models} shows that the cloud has nearly constant temperature of 20 K in model A.
Since FUV radiation is not significant in this model, effective cooling keeps the dense region relatively cold.  

Comparisons of model B to model A show that FUV radiation has little influence on the density and the thermal structures of the cloud.
Although the cloud surface is heated near 30 K in model B, this warm surface layer is very thin.
High density at the head region prevents CO molecules, which are main coolant of molecular gas, from 
photodissociation except thin surface layer. 
Isothermal gas is good approximation at inner region of the cloud.
Figure \ref{fig column density profiles} shows that the column density of
model B is much higher than observations as well as model A.
Although FUV radiation reduces the density than model A by factor of a few, it
is not enough to reproduce observed low column densities.
The shape of the cloud slightly differs from model A. Strong FUV radiation enhances thermal pressure at head region and makes 
the curvature radius at cloud tip larger than that of model A. Moving to tail side, the differences of the density and the column density from those 
of model A become larger. As a result, the slope of the column density profile of model B is steeper 
than that of model A.  

Comparisons of model E2 to model A and model B show that magnetic field reduces the density of the cloud.
Figure \ref{fig comparison three models} shows that the density and column density
of model E2 at head region are one
order of magnitude lower than those of model A and B. 
As shown in \ref{subsec alpha}, magnetic pressure is dominant at $z > -0.4$ pc in model E2.
Additional support due to magnetic pressure makes density of the cloud lower than models without magnetic field.
The warm surface layer of model E2 is thicker than that of model B, because lower density at head region allows FUV radiation penetrating 
deeper inside the cloud.
Figure \ref{fig column density profiles} shows that the slope of the column density profile of model E2 is flatter 
than those of model A and B.  
The column density profile of model E2 shows better agreement with that of the
observed cloud than other two models.

\subsection{Dependence on the Density Dependence of Magnetic Field Strength}\label{subsec alpha}
In this subsection, we present comparison of following three models, C2, D2, and E2 to show how the value of exponent $\alpha$ 
in equation (\ref{dependence of magnetic field}) affects the results. 
The values of $\alpha$ and $B_0$ in model C2, D2, and E2 were set to be 0.25 and 45 $\mu$G, 0.50 and 15 $\mu$G, and 0.75 
and 5 $\mu$G, respectively. 
Comparison of these models reveals that the value of $\alpha$ influences structure of cloud. 
 
As can be seen from Figure \ref{fig column density profiles}, although these three models have similar maximum values of column 
density of $\sim 2 \times 10^{22}$ $\mathrm{cm^{-2}}$, 
the column density profiles of these models are qualitatively different. 
The model with smaller value of $\alpha$ has steeper column density profile at the head region (z $>$ -0.2 pc)  
and has flatter column density profile at tail region (z $<$ -0.3 pc).
In model C2, the slope of the column density profile becomes flatter with moving to tail side.
Opposite trend is observed in model E2.
The slope of the column density profile becomes steeper with moving to tail side.
Contrary to model C2 and E2, the slope of the column density profile is nearly constant through entire region in model D2. 

\begin{figure*}
  \begin{center}

  \epsscale{0.9}
  \plotone{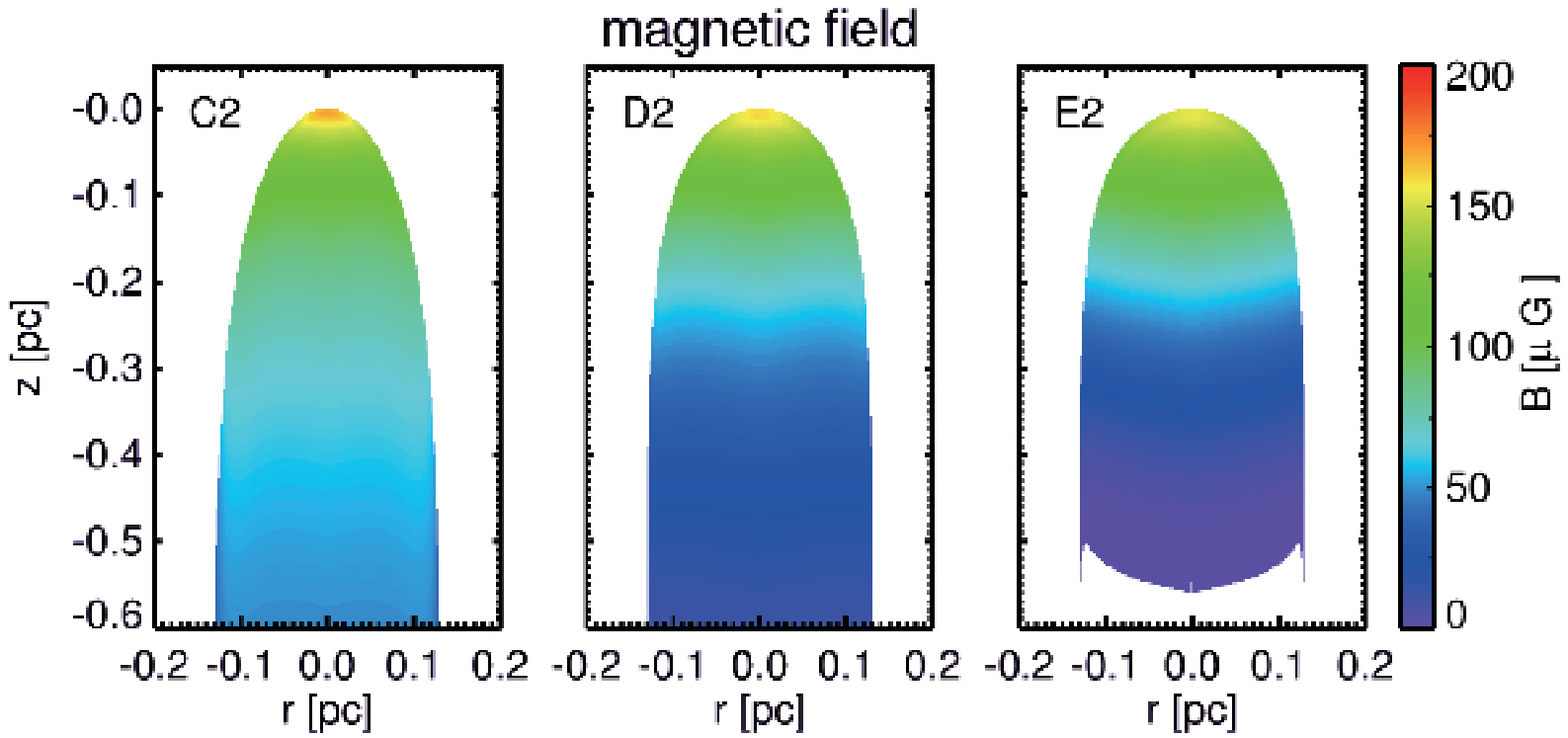}\\
  \plotone{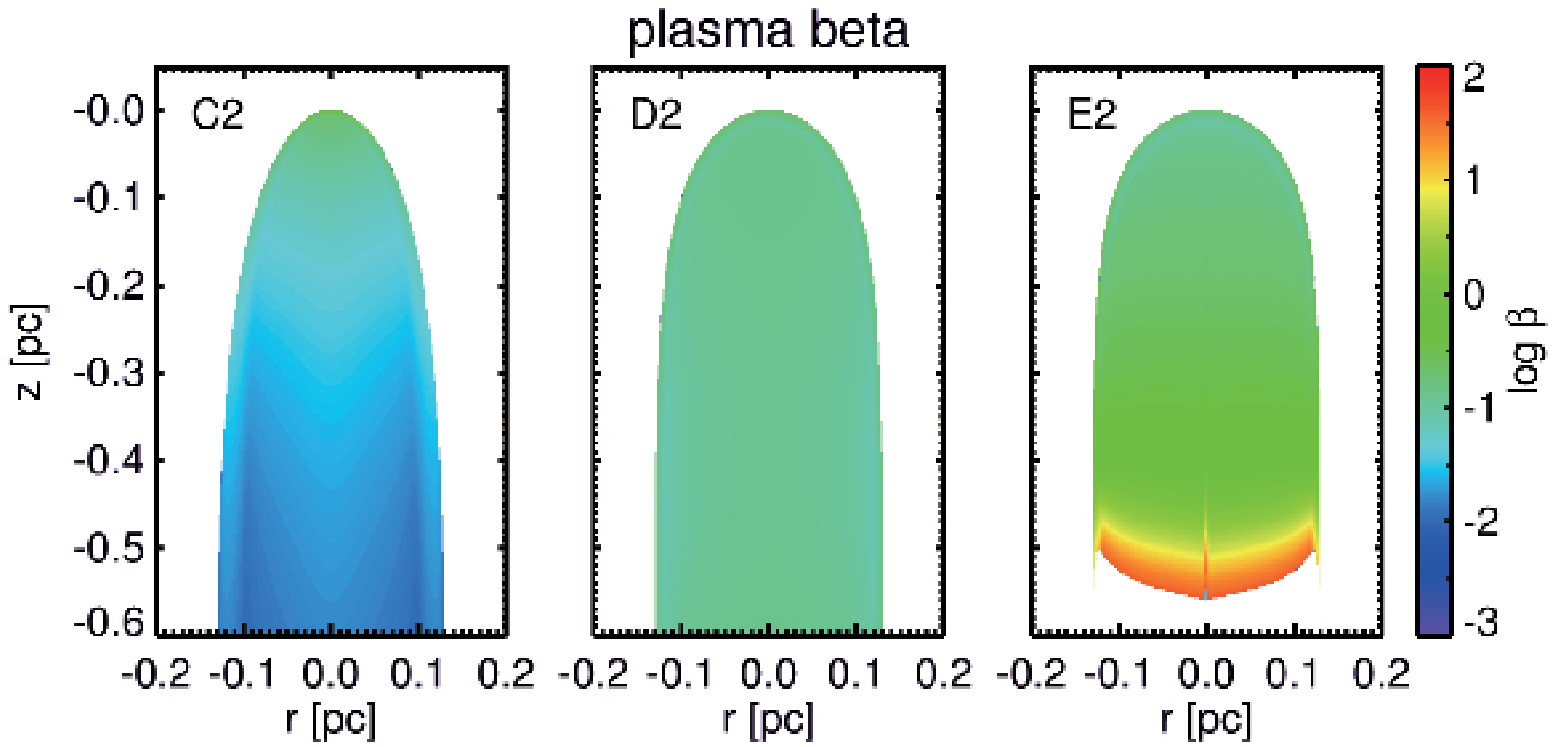}

  \end{center}
  \caption{The magnetic field strength and the ratio of thermal pressure to the magnetic pressure 
        in model C2 (left), D2 (center), and E2 (right).}
  \label{fig strength and beta}
 \end{figure*}

Figure \ref{fig strength and beta} shows the magnetic field strength and plasma beta, which is defined as the ratio of thermal pressure 
to magnetic pressure, in these models.
Although these three models have the similar magnetic field strengths of $\sim$ 150 $\mathrm{\mu G}$ at cloud tip,  
the magnetic field strength decreases more quickly with moving to tail side in the model with larger value of $\alpha$. 
The distribution of plasma beta in the cloud qualitatively changes whether $\alpha$ exceeds 0.5 or not.
As can be seen from equation (\ref{total pressure}) and (\ref{dependence of magnetic field}), the thermal pressure and the magnetic 
pressure are proportional to $\rho$ and $\rho^{2 \alpha}$, respectively.
When $\alpha$ is larger than 0.5, magnetic pressure increase faster than thermal pressure with increasing the density, 
and magnetic pressure become dominant at dense region. 
When $\alpha$ is smaller than 0.5, by contrast, thermal pressure become dominant at dense region. 
In model C2 in which $\alpha = 0.25$, the plasma beta is larger than unity at
dense head region and decreases with moving to tail side.
Thus, the thermal pressure is dominant at the head region; the magnetic pressure is dominant at tail side.
In model E2 in which $\alpha = 0.75$, opposite trend is observed.
The magnetic pressure is dominant at the head region; the thermal pressure is dominant at tail side.
In model D2 in which $\alpha=0.5$, the plasma beta is nearly constant at entire cloud, because the magnetic pressure and the thermal 
pressure increase at the same rate with increasing of the density.

\begin{table*}
\begin{center}
\caption{Summary of numerical results\label{tbl results}}
\begin{tabular}{ccccccc}
\tableline
\tableline
Model & $R_c$ [pc] & $T_{tip}\tablenotemark{a}$ [K] &  $n_{\mathrm{H_2},max}\tablenotemark{b}$ [$\mathrm{cm^{-3}}$] &  
$N_{\mathrm{H_2},max}\tablenotemark{c}$  [$\mathrm{cm^{-2}}$] &  
$B_{max}$\tablenotemark{d} [$\mu$G]   &  $M_{cl}$\tablenotemark{e}  [M$_{\sun}$]  \\
\tableline

A   & $7.6 \times 10^{-2}$ &  19.5    &  $4.48 \times 10^{5}$ & $1.26 \times 10^{23}$ &  -  &  61.3  \\
B   & $9.4 \times 10^{-2}$ &  28.9    &  $4.19 \times 10^{5}$ & $1.27 \times 10^{23}$ &  -  &  54.8  \\
C1  & $7.0 \times 10^{-2}$ &  30.3    &  $2.61 \times 10^{5}$ & $5.65 \times 10^{22}$ & 119 &  24.1  \\
C2  & $4.1 \times 10^{-2}$ &  29.5    &  $1.07 \times 10^{5}$ & $1.61 \times 10^{22}$ & 172 &  5.59  \\
C3  & $3.2 \times 10^{-2}$ &  27.2    &  $5.35 \times 10^{4}$ & $6.36 \times 10^{21}$ & 192 &  2.03  \\
C4  & $2.7 \times 10^{-2}$ &  24.1    &  $2.27 \times 10^{4}$ & $2.15 \times 10^{21}$ & 208 &  0.653 \\
D1  & $8.2 \times 10^{-2}$ &  30.3    &  $1.81 \times 10^{5}$ & $5.33 \times 10^{22}$ & 133 &  27.7  \\
D2  & $7.1 \times 10^{-2}$ &  27.5    &  $5.75 \times 10^{4}$ & $1.81 \times 10^{22}$ & 161 &  10.4  \\
D3  & $6.7 \times 10^{-2}$ &  24.2    &  $2.39 \times 10^{4}$ & $7.17 \times 10^{21}$ & 173 &  4.20  \\
D4  & $6.5 \times 10^{-2}$ &  21.2    &  $7.83 \times 10^{3}$ & $2.27 \times 10^{21}$ & 178 &  1.35  \\
E1  & $9.0 \times 10^{-2}$ &  30.1    &  $1.42 \times 10^{5}$ & $5.17 \times 10^{22}$ & 139 &  28.4  \\
E2  & $8.9 \times 10^{-2}$ &  26.9    &  $4.87 \times 10^{4}$ & $2.05 \times 10^{22}$ & 155 &  13.0  \\
E3  & $8.8 \times 10^{-2}$ &  23.8    &  $2.08 \times 10^{4}$ & $8.77 \times 10^{21}$ & 164 &  5.81  \\
E4  & $8.7 \times 10^{-2}$ &  20.9    &  $6.35 \times 10^{3}$ & $2.65 \times 10^{21}$ & 168 &  1.79  \\
\tableline
\end{tabular}
\tablenotetext{1}{Temperature at cloud tip}
\tablenotetext{2}{Maximum value of number density of H$_2$}
\tablenotetext{3}{Maximum value of column density of H$_2$}
\tablenotetext{4}{Maximum value of magnetic field strength}
\tablenotetext{5}{Total cloud mass}
\end{center}
\end{table*}

\subsection{Dependence on Initial Magnetic Field Strength}\label{sect B_0}
\begin{figure}
  \begin{center} 

  \epsscale{1.0}
  \plotone{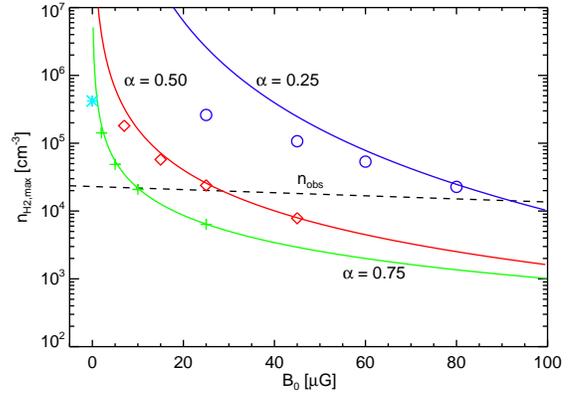}

  \caption{The maximum number densities as function of initial magnetic field
      strength. The circles (colored blue in the online version), the diamonds
      (colored red in the online version), the pluses (colored green in the online
      version), and the asterisk (colored light blue in the online version)
      represent model C1-C4, model D1-D4, model E1-E4, and model B, respectively. The solid lines represent 
     analytic estimates given by equation (\ref{eq n_tip}). \newline 
     (A color version of this figure is available in the online journal.)}
 \label{fig dependence of nmax}
  \end{center}
 \end{figure}

\begin{figure}
  \begin{center} 

  \plotone{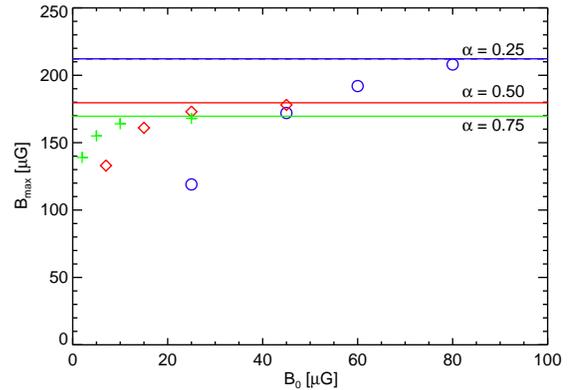}

  \caption{The maximum magnetic field strengths as function of initial
      magnetic field strength. The circles  (colored blue in the online version), the diamonds (colored red in the online version), 
      and the pluses (colored green in the online version) represent 
      model C1-C4, model D1-D4, and model E1-E4, respectively. The solid lines represent analytic asymptotic values given by equation 
     (\ref{eq b_tip}). \newline 
     (A color version of this figure is available in the online journal.)}
 \label{fig dependence of bmax}
  \end{center}
 \end{figure}

In this subsection, we describe general trends which arise when initial magnetic field strength $B_0$ is changed.
In Figure \ref{fig dependence of nmax} and \ref{fig dependence of bmax}, maximum values of the number density 
and the magnetic field strength are plotted as function of initial magnetic field strength $B_0$, respectively.
As shown in Figure \ref{fig comparison three models} and \ref{fig strength and beta}, the number density and the magnetic field 
strength reach their maximum values near cloud tip, when effects of FUV radiation are not significant.
We focus on low plasma regime and consider these maximum values as values at cloud tips.

We can obtain asymptotic behavior of number density in low plasma beta regime
by neglecting thermal pressure.
From equation (\ref{total pressure}) and (\ref{dependence of magnetic field}), the number density is expressed as
\begin{equation}
   n = n_0 \left( 8 \pi P_{tot} \right)^{1 / 2 \alpha} B_0^{-1 / \alpha},
\end{equation}
where we neglect the first term in the right-hand side of equation (\ref{total pressure}), because we assume that 
the magnetic pressure is dominant. 
Total pressure $P_{tot}$ at cloud tip is given by substituting $\theta = 0$ into equation (\ref{surface pressure}).  
Therefore, using the relation of $F_{UV} = F_i / q$, we obtain the number density at cloud tip as
\begin{equation}
    n_{tip} = n_0 \left( \frac{16 \pi  \mu m_{\rm H} c_i F_i}{q}  \right)^{1 / 2 \alpha} B_0^{-1 / \alpha}.
  \label{eq n_tip}
\end{equation}
This relation shows that the density at cloud tip is proportional to $B_0^{-1 / \alpha}$ in low plasma beta regime. 
The maximum number densities of our numerical models approach lines representing 
this analytic asymptotic value as $B_0$ increases (see Figure \ref{fig dependence of nmax}). 
We adopted minimum value of $R_c$ for models with each $\alpha$ to calculate $q$ in equation (\ref{eq n_tip}), 
because curvature radius at cloud tip $R_c$ decreases with increasing of $B_0$.  
We used the value of $R_c = 0.027$ pc for models with $\alpha = 0.25$,  $R_c = 0.065$ pc for models with $\alpha = 0.50$, and 
$R_c = 0.087$ pc for models with $\alpha = 0.75$. 
The estimate value of maximum number density of SFO 22 is also plotted in this figure.
Dividing maximum value of observed column density $1.89 \times 10^{22} \, \mathrm{cm^{-2}}$ by cloud width of 
0.26 pc gives rough estimate value of $n_{obs}=2.36 \times 10^4 \, \mathrm{cm^{-3}}$. 
From this figure, it is inferred that SFO 22 had initial magnetic field strength of  
several to $\sim$90 $\mu$G, if value of $\alpha$ is from 0.25 to 0.75.  

We can also obtain asymptotic behavior of magnetic field strength in low plasma beta regime.
Substituting equation (\ref{eq number density}) and (\ref{eq n_tip}) into equation (\ref{dependence of magnetic field}) gives the magnetic field 
strength at cloud tip as
\begin{equation}
    B_{tip}  =  \left( \frac{16 \pi  \mu m_{\rm H} c_i F_i}{q}  \right)^{1 / 2}.  
  \label{eq b_tip}
\end{equation}
This relation shows that magnetic field strength at cloud tip is function of the curvature radius of the cloud at tip $R_c$ 
and incident ionizing photon flux $F_i$, 
but independent of the initial magnetic field strength $B_0$ and initial
number density $n_0$. 
As the initial magnetic field strength $B_0$ increases, the magnetic field strength at cloud tip approaches this asymptotic value.
The equation (\ref{eq b_tip}) gives asymptotic value of 212 $\mu$G for models with $\alpha = 0.25$, 180 $\mu$G for models with $\alpha = 0.50$, 
and 170 $\mu$G for models with $\alpha = 0.75$. 
The maximum magnetic field strengths of our numerical models asymptotically approach 
this value at large $B_0$ (see Figure \ref{fig dependence of bmax}). 

\section{discussion}\label{sect discussion}

\subsection{Comparison between Observations and Numerical Models}\label{sect comparison}
We developed numerical model for photo-evaporating cloud which is in quasi-stationary equilibrium state, assuming that pressure 
gradient of the cloud balances the inertia force caused by acceleration due to the back 
reaction of photoevaporation flow.
As shown in \ref{sect FUV and B}, observed density structure of BRC SFO 22 can not be explained 
by the reference model (model A) in which either effects of strong FUV radiation
or effects of magnetic field were not included.
The column density of the reference model is one order of magnitude higher than that of observed cloud.
Observed lower column density implies that the pressure of observed cloud is
higher than that of the reference model. 
There are two possible explanations for this large difference of density structures between observed cloud and 
the reference model.
The one explanation is that heating due to strong FUV radiation from exciting star warms the cloud, hence 
enhanced thermal pressure makes the cloud reach equilibrium state with lower density than the reference model.
Another explanation is that additional pressure due to magnetic field makes the cloud reach equilibrium state 
with lower density than the reference model.

Comparisons between the reference model and numerical model including effects of strong FUV radiation (model B) show that strong 
FUV radiation has little influence on structure of the cloud. 
Although strong FUV radiation slightly reduces the density and the column density of the cloud, the column density is much 
higher than that of observed cloud.
Small difference from the reference model suggests that heating due to FUV does
not affect structure of the cloud at least in SFO 22.
To affect on structure of the cloud, the photoelectric heating needs to be dominant heating process, {\it i.e.} $\Gamma_{pe} / \Gamma_{cr} > 1$.  
For rough estimation, let us approximate the efficiency of photoelectric heating $\epsilon$ expressed by equation 
(\ref{eq efficiency pe}) by a constant value of $4.87 \times 10^{-2}$. 
From equation (\ref{eq heat pe}), (\ref{eq heat cr}), and the relation of $\tau_{UV} = 2.5 A_V$, the ratio of photoelectric 
heating rate to cosmic ray heating rate is written as 
\begin{equation}
    \frac{\Gamma_{pe}}{\Gamma_{cr}} \simeq 43 \, G_0 \exp(- 2.5 A_v).
\end{equation}
Therefore, visual extinction of the cloud along the direction to exciting star is need to be smaller than critical value
\begin{equation}
    A_{V, cri} = \frac{\ln (43 G_0)}{2.5},
\end{equation}
Applying this equation for SFO 22, it gives the critical value $A_{V,cri}$ as 3.3. 
The total mass traced by $^{13}$CO emission and width of SFO 22 give rough estimate of visual extinction of SFO 22 as
\begin{equation}
    A_{V,SFO22} \simeq  X_{A_{V}} \frac{12.0 \, \mathrm{M_{\sun}}}{\mu m_{\rm H} \pi (0.13 \, \mathrm{pc})^2 },
\end{equation}
It gives the value of 15.4. This larger value of $A_{V,SFO22}$ compared to $A_{V, cri}$ indicates that the heating due to 
FUV radiation affects only near cloud surface. 
This discussion based on rough estimation is consistent with our numerical results.

Comparisons between the reference model and numerical model including effects of magnetic field show that magnetic field 
strongly affects structure of the cloud.
From results of model C2, D2, and E2, initial magnetic field strength of 5 - 45 $\mu$G is required to reproduce observed column 
density of $\sim 2.0 \times 10^{22}$ $\mathrm{cm^{-2}}$ depending on the value of $\alpha$.
On the other hand, magnetic field strengths in molecular clouds are measured by Zeeman effects.
\cite{2010ApJ...725..466C} statistically analyzed samples of clouds with Zeeman observation in order to infer the distribution of 
the total magnetic field strength in the samples.
According to their analysis, molecular clouds with the density of $\sim 10^3$ $\mathrm{cm^{-3}}$ have magnetic field strength of a 
few to several tens of $\mathrm{\mu G}$. This coincidence of magnetic field strengths between numerical models and observations shows  
that observed column density of SFO 22 are naturally explained by effects of magnetic field.

\subsection{Magnetic Field Configurations in BRCs}\label{sect B field of sfo22}
\begin{figure}
  \begin{center}
    \epsscale{0.9}
    \plotone{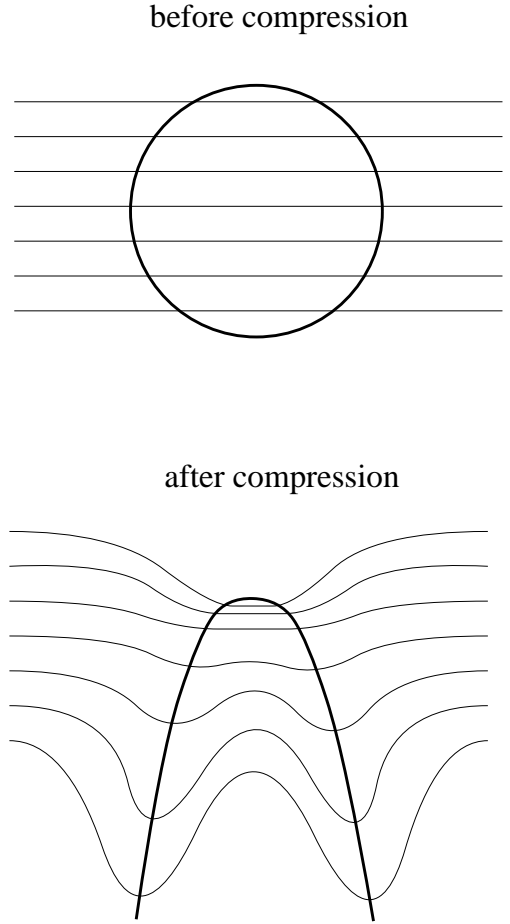}
  \end{center}
       \caption{Schematic figures of magnetic field configuration in SFO 22 before undergoing compression by radiation-driven implosion (top) 
       and after reaching the phase of quasi-stationary equilibrium (bottom).  }
     \label{figure B field}
\end{figure}

As shown in \ref{subsec alpha}, structure of BRC strongly depends on the
value of $\alpha$ in equation (\ref{dependence of magnetic field}). 
The slope of column density profile at head region becomes steeper with decreasing $\alpha$.
The slope of column density profile at tail region shows opposite dependence on $\alpha$.
Since this parameter represents how much magnetic field is trapped in the gas during the compression, 
our numerical results implies that direction of magnetic field affects evolution of BRCs. 
When magnetic field is parallel to the direction of UV radiation, the value of $\alpha$ is close to 0. 
On the other hand, when magnetic field is perpendicular to the direction of UV radiation, the value of $\alpha$ is close to 1. 
Since results of model E2 in which $\alpha$ was set to be 0.75 shows best agreement with
observations, magnetic field in SFO 22 is thought to be near 
perpendicular to UV radiation. 
However, deviation from observations can be seen at tail region. The slope of column density of model E2 is steeper than that of
observed cloud at tail region (see Figure \ref{fig column density profiles}). 
This difference of column density profiles implies possibility that the value of $\alpha$ at tail region is smaller than head region.   
\cite{2011MNRAS.412.2079M} performed three dimensional MHD simulations and found that for weak and medium magnetic field 
strengths an initially perpendicular field is swept into alignment with tail during dynamical evolution.
Their results may explain the reason why the value of $\alpha$ is small at tail region.  
Figure \ref{figure B field} shows schematic figure of magnetic field configuration in SFO 22 suggested by comparisons between our 
numerical results and observations. 
The possible scenario is as follows. 
Magnetic field in SFO 22 was initially perpendicular to UV radiation from exciting star. 
Then, compression due to radiation-driven implosion make magnetic field be close to parallel to UV radiation at tail region. 

Very few attempts have been made at observational study on magnetic field of BRCs.
\cite{1996MNRAS.279.1191S} performed optical polarimetry observations toward CG 22 and reported that magnetic field is parallel to its tail.
Their rough estimate gives magnetic field strength of $\sim 30 \mathrm{\mu G}$. 
It is comparable with magnetic field strength obtained by our numerical models (see Figure \ref{fig strength and beta}). 
Other optical polarimetry observations by \cite{1999MNRAS.308...40B} showed that magnetic field in CG 30-31 is found to be nearly 
perpendicular to the cometary tails.
Observational studies on relations between density structures and
magnetic field configurations in BRCs are required to reveal effects of 
magnetic field on evolutions of BRCs.


\section{conclusions}\label{sect conclusions}
Using the Nobeyama 45 m telescope, we observed BRC SFO 22 in the $^{12}$CO (J = 1-0), $^{13}$CO (J = 1-0), and C$^{18}$O (J = 1-0) lines. 
Observed column density profiles were compared with those of numerical models for photo-evaporating cloud in 
quasi-stationary equilibrium state in order to investigate how magnetic field
and heating due to strong FUV radiation from exciting star affect structures of BRCs. 
We summarize our main conclusions as follows:

\begin{enumerate}
   \item From our radio observations, the column density profiles of SFO 22
       along the line to its exciting star are nearly flat with 
       the column density of $\sim 10^{22}$ $\mathrm{cm^{-2}}$.
   
   \item Strong FUV radiation from exciting star has little influence on structure of 
         SFO 22. 
         Although enhanced thermal pressure due to strong FUV radiation slightly reduces the density of the cloud, its effects are 
         not enough to reproduce the observed density structure of SFO 22.

   \item Magnetic field strength and direction of magnetic field strongly affect structures of BRCs.   
         Numerical model with initial magnetic field strength of $5 ~ \mathrm{\mu G}$
         shows the best agreement with the observations. 
         When magnetic field is nearly parallel to UV radiation from exciting
         star, the cloud has steep column density profile at head region  
         and flat column density profile at tail region.
         When magnetic field is nearly perpendicular to UV radiation, the column density profile shows opposite trend.

\end{enumerate}

In this paper we only focus on quasi-stationary equilibrium phase of radiation-driven implosion model, and we will discuss implosion phase 
in a subsequent paper.
We also plan further study using MHD simulation to establish more realistic evolutionary model of BRCs.

\acknowledgments
The authors would like to thank Kohji Sugitani, Shin-ya Nitta, and 
Hiroyuki Takahashi for helpful discussions. 
We are grateful to the staff of Nobeyama Radio Observatory for their help 
during the observations. 
This work is supported by the Theoretical Institute for Advanced Research
in Astrophysics (TIARA) operated under Academia Sinica in Taiwan.
Numerical computations were in part carried out on the general-purpose PC 
farm at Center for Computational Astrophysics, CfCA, of National 
Astronomical Observatory of Japan. 
Numerical computations were carried out by using a workflow system, RENKEI-WFT, 
which is developed by National Institutes of Informatics, Japan.


\clearpage

\end{document}